\documentclass[12pt]{iopart}
\usepackage{xcolor}
\usepackage{iopams}
\usepackage{mathrsfs}
\usepackage{latexsym}
\usepackage{graphicx}
\usepackage{subcaption}
\usepackage{hyperref}
\hypersetup{
    colorlinks=true
    }
\usepackage{ulem}
\usepackage{tabularx}
\usepackage[T1]{fontenc}
\usepackage[utf8]{inputenc}
\usepackage{diagbox}
\usepackage{listings}

\newcommand{\code}[1]{\texttt{#1}}

\newcommand{\text}[1]{{\rm #1}}

\newcommand{\cactus}[0]{\code{Cactus}}
\newcommand{\ilgrmhd}{\code{IGM} }
\newcommand{\etk}{\code{Einstein Toolkit} }
\newcommand{\wyodoi}{\href{https://doi.org/10.15786/20452374}{doi:10.15786/20452374}}
\newcommand{\ilgrmhdpaper}[1][~]{the ILGRMHD paper#1}
\newcommand{\ilgrmhdresult}{IGM paper results}

\begin{document}

\title[HPC-driven computational reproducibility: IllinoisGRMHD]{HPC-driven computational reproducibility in numerical relativity codes: A use case study with IllinoisGRMHD}

\author{Yufeng Luo$^{1,2,3,4,6}$, Qian Zhang$^5$, Roland Haas$^{4,3}$, Zachariah B. Etienne$^7$,
  Gabrielle Allen$^{2,1,4}$}
\address{$^1$ Department of Physics and Astronomy, University of Wyoming, Laramie, Wyoming, 82071, USA}
\address{$^2$ School of Computing, University of Wyoming, Laramie, Wyoming, 82071, USA}
\address{$^3$ Department of Physics, University of Illinois at Urbana-Champaign, Urbana, Illinois, 61801, USA}
\address{$^4$ NCSA, University of Illinois at Urbana-Champaign, Urbana, Illinois, 61801, USA}
\address{$^5$ Strategy \& Planning, Digital Research Alliance of Canada, Toronto, Ontario  M4S 3C6 Canada}
\address{$^6$ Department of Astronomy, University of Illinois at Urbana-Champaign, Urbana, Illinois, 61801, USA}
\address{$^7$ Department of Physics, University of Idaho, Moscow, ID 83844, USA}

\ead{yluo4@uwyo.edu}

\begin{abstract}
Reproducibility of results is a cornerstone of the scientific method.
Scientific computing encounters two challenges when aiming for this goal.
Firstly, reproducibility should not depend on details of the runtime environment, such as the compiler version or computing environment, so results are verifiable by third-parties. Secondly, different versions of software code executed in the same runtime environment should produce \textbf{consistent numerical results for physical quantities}.
In this manuscript, we test the feasibility of reproducing scientific results
obtained using the \code{IllinoisGRMHD} code that is part of an open-source community software for simulation in relativistic astrophysics, the 
\code{Einstein Toolkit}. We verify that numerical results of simulating a single isolated
neutron star with \code{IllinoisGRMHD} can be reproduced, and compare them to results reported by the code authors in 2015. We use two different supercomputers: Expanse at SDSC, and Stampede2 at
TACC. 

By compiling the source code archived along with the paper on both Expanse and
Stampede2, we find that \code{IllinoisGRMHD} reproduces results published in its
announcement paper up to errors comparable to round-off level changes in initial
data parameters. We also verify that a current version of
\code{IllinoisGRMHD} reproduces these results once we account for bug fixes
which has occurred since the original publication.

\end{abstract}

%
\vspace{2pc}

\section{Introduction: computational reproducibility}\label{sec:intro}

\subsection{Defining computational reproducibility}
Computational research, or scientific computing, uses advanced research computing capabilities to understand and solve complex problems in science. Computational research spans many disciplines, but at its core, it involves the development and implementation of mathematical models and numerical simulations applied to data. The main purpose of reproducibility is to verify the scientific method and outputs, and provide a mechanism to confirm or refute a study's conclusions. That is why reproducibility is a Process, not an Achievement~\cite{Lin_Zhang_2020}. In this study,  HPC-driven computational reproducibility has a loose definition, which is to obtain consistent scientific outputs rather than exact results using the original artifacts. This reproducibility experiment is conducted by a different team with the same experimental setup, including
the same input data, the same numerical model, and following the same method and computational steps, but using a different computational environment (HPC cluster).
Due to differences in compilers and hardware, some machine- and environment-specific parts of the source code, mainly configuration files need to be modified so that it can be compiled and run on a new cluster.

\subsection{Computational Reproducibility and FAIR principles}\label{sec:fair}
The FAIR (Findable, Accessible, Interoperable, and Reusable) data principles~\cite{wilkinson2016fair}, aim to enhance and support the reuse of digital material by both humans and machines.
A high-level principle of computational reproducibility is to provide a clear, specific, and complete description of how a reported result was reached, although different areas of study or types of inquiry may require different kinds of information.
Although scientific software differs from research data, the high-level FAIR data principles also apply to software code, in terms of the goals that ensure and improve the findability, accessibility, interoperability, reusability, transparency and optimal use of research objects.
A computationally reproducible research package may include data (primary and secondary data), software program(s) and documentation (including software dependencies and runtime / computational environment) for replicating published results, and capturing related provenance information, etc.
Over the last few years, a number of groups have been working towards the development of a set of FAIR guiding principles for research software (RS), including the FAIR For Research Software Working Group (FAIR4RS WG)~\cite{FAIR4RSWG:web,FAIR4RS_2020} which is co-led by RDA~\cite{rda:web}, FORCE11~\cite{force11:web}, and the efforts of Research Software Alliance (ReSA)~\cite{ReSA:web}, the Software Sustainability Institute (SSI)~\cite{SSI:web} and grassroots communities (e.g., UK Reproducibility Network~\cite{ukrn:web}).

\subsection{Computational reproducibility challenges}
Although research reproducibility is a critical and continuous component of the 
scholarly communications process, computational irreproducibility cannot be 
traced to one single cause. From the research software (RS) perspective, there 
are multiple factors that contribute to the lack of reproducibility: RS is not 
widely disseminated or shared and not readily discoverable and thus 
inaccessible, inhibiting research transparency, reproducibility and 
verification. As one of the steps toward scientific reproducibility, RS should 
be properly cited so that it is uniquely identified (e.g., the specific version 
of any RS package that is used to produce respective results), which also 
benefits transparency and traceability of research results. The Accessibility 
principle of the FORCE11 Software Citation Principles states that ``software 
citations should permit and facilitate access to the software itself and to its 
associated metadata, documentation, data, and other materials necessary for 
both humans and machines to make informed use of the referenced software.'' 
While this does not require that the RS be freely available, the metadata 
should be, and should provide sufficient information for the RS to be accessed 
and used. The development, deployment, and maintenance of reusable RS (whether 
computational in nature, or that relies on any software-based 
analysis/interpretation) are increasingly  recognized internationally as a key 
part of facilitating trusted, reproducible research outputs and open science. 
Software versioning, a robust testing/quality framework (e.g. verification and 
validation), code repositories, and portability, all of which are recognized as 
desirable aspects of software quality, have all helped to drive the rapid 
evolution of research reproducibility. Software sustainability is key to 
reproducible science too, as it provides a critical tool for effective 
review and analysis of published results, which may lead to new research 
efforts. However, the wide range of robust frameworks and approaches for 
curating and preserving RS as a complex digital object represents a significant 
challenge for sustainable access, thereby hindering research reproducibility.

At the cultural and societal level, transformation to an open science-driven RS culture depends on the creation of tools, platforms and services that enable researchers to mobilize knowledge and make research processes more efficient, transparent, reproducible, and responsive to societal challenges. Specific elements of this shift include: increasing collaboration and interaction among researchers; the development of technical infrastructure that promotes the adoption of emerging research practices; the development, promotion, and adoption of open-source and open-science practices. These shifts require an agile and responsive ecosystem with strong RS workforce support and sustainable funding.

\subsection{Computational reproducibility with the Einstein Toolkit}

The historical lack of direct observational data in numerical relativity targeting
compact objects coalescence led to simulation and the development of robust and reliable
software codes being a primary scientific approach in this field.
Numerical relativity (NR) 
is a discipline that combines general relativity with numerical simulations to 
study the physics of compact objects, such as binary neutron stars and black 
holes. NR transforms theoretical models for a system into executable codes and 
simulates the system using the codes to produce physical observables, such as 
gravitational waves, that can be detected and verified by experiments or 
astronomy observations, e.g. the Laser Interferometer Gravitational-Wave 
Observatory (LIGO). Numerical relativity codes have been crucial in the development
of the gravitational wave templates used to understand the physical parameters of
the binary black hole and binary neutron star systems detected by the LIGO-Virgo
Collaboration~\cite{PhysRevLett.116.061102}. With 
both open source and reproducibility being considered important aspects of 
numerical simulations, from among a selection of current open- and closed-source
astrophysics 
codes GRChombo~\cite{Andrade2021}, SpECTRE~\cite{spectrecode}, SpEC~\cite{speccode}, 
DendroGR~\cite{doi:10.1137/18M1196972} and BAM~\cite{Bruegmann:2003aw}
we use the Einstein Toolkit to perform our experiments due to its wide use and 
support of many different computing clusters.

The reproducibility experiment described here is based on a use case for the Einstein Toolkit. 
The Einstein Toolkit is an open-source, community-driven cyberinfrastructure
ecosystem that provides key computational tools to support research in computational astrophysics, gravitational physics,
and fundamental science. The Einstein Toolkit community involves experts with
diverse backgrounds, from gravitational physics and astronomy to computer
science and engineering. As such, the Einstein Toolkit evolves and grows---just as
fundamental science itself progresses---to facilitate novel applications with
ambitious science goals and high productivity of its users, and to respond to the
needs of new community members.

The Einstein Toolkit is built on the \cactus~\cite{Cactuscode:web} computational framework to connect different
modules and to achieve a clean separation between science and infrastructure
components. This enables domain experts in astrophysics and computer engineering
to focus their efforts on the components they are most comfortable dealing with.
All components within the Einstein Toolkit are distributed using free and open-source licenses enabling users to mix and match modules, adapt modules to their
own needs and share these modules freely with collaborators. This arrangement, while flexible and allowing for easy collaboration among distributed and non-coordinating groups, poses both opportunities and challenges with respect to
ensuring reproducible simulations.

It is worth mentioning that the numerical simulation framework --- \cactus~\cite{Cactuscode:web} already puts a premium on reproducibility and portability.
In particular, the \code{Cactus} framework includes basic infrastructure to ensure the reproducibility of results as the code evolves, via
its included test suite mechanism. A set of system
level regression test suites consist of input descriptions for \code{Cactus}
in the form of parameter files as well as expected output files and an error
threshold value, which are provided by the code authors. \cactus' infrastructure lets developers and users re-run these
test suites and verify that the current code passes all test suites, and ensures
 all code changes that result in changes in data beyond the test suite
threshold value are detected, based on which the developers can choose to either update the test data or fix the newly
introduced bug. 

\cactus\ also contains a module, ``\code{Formaline}'', that collects all
source files used to compile the simulation executable and embeds an archive of
these files in the executable itself. In addition, \code{Formaline} generates a
unique identifier for each simulation executable and each simulation run.
At run time the
executable outputs a copy of the included archive files along with its regular
simulation output. Each output file is also tagged with the unique identifier 
of the simulation executable and simulation run. This way all code used to 
generate a set of
output files are included, alongside those files and all output files record the exact
code used to produce them. 

Together \code{Formaline} and the test-suites provide mechanisms ensuring
reproducible simulations by recording the code version used to produce results and 
tracking code changes that affect results.

The computational reproducibility experiment described in this paper follows the current practices of FAIR principles for data and RS, respectively. The raw simulation results and analysis code are findable and accessible through the WyoScholar data depository with \wyodoi. The figures in this paper are reproducible with the containerized environment included in the analysis code with Docker.

\section{Use case study}

As a concrete example of the challenges faced in achieving computational
reproducibility in HPC computations, we reproduce results obtained using the \code{IllinoisGRMHD}
code~\cite{Etienne:2015cea} which was first officially included in the \code{ET\_2015\_11} ``Somerville'' release of the \code{Einstein Toolkit}.
In the manuscript announcing \code{IllinoisGRMHD}~\cite{Etienne:2015cea} the authors evolved solutions for a TOV star in general relativistic hydrodynamics 
and compared their results to those obtained by other codes. The TOV star is a spherically symmetric, nonrotating neutron star assuming an equation of state that represents initially cold, degenerate nuclear matter. This TOV star does not have a magnetic field. 
The single polytropic EOS for the TOV star is $P=K\rho^{\Gamma}$, where $K=1$, $\Gamma=2$. This is the same setup as in Appendix A of \cite{Etienne:2015cea}.
Our aim is to reproduce 
the results described in that paper. In the following text, we refer to \code{IllinoisGRMHD} manuscript \cite{Etienne:2015cea} as \ilgrmhdpaper[], and we refer to the \ilgrmhdpaper[] results as \ilgrmhdresult.

We perform the case study
on two different
supercomputers,
SDSC Expanse~\cite{expanse:web} and TACC Stampede2~\cite{stampede2:web}, each evolving
the same dataset constituting the initial condition that was used in \ilgrmhdpaper
and we compared our results to \ilgrmhdresult. In order to differentiate between
changes due to modifications to simulation code and changes due to differences in the supercomputer environment,  we used two versions of the \code{IllinoisGRMHD}: 
\begin{itemize}
    \item the most recent \code{IllinoisGRMHD} from 
    the \code{ET\_2022\_11} ``Sophie Kowalevski'' release of the \code{Einstein Toolkit}, called \ilgrmhd 2022 in the following
    \item the original \code{IllinoisGRMHD} used in~\cite{Etienne:2015cea}, dubbed \ilgrmhd 2015 available in~\cite{zachariah_etienne_2022_7545717}.
\end{itemize}
\code{IllinoisGRMHD}'s
complete code history
is available in its public source code repository~\cite{IllinoisGRMHD:git}  using git, which we used
to track down commits introducing any observed change in behavior. \ilgrmhd 2015 can be found on the original publication authors' website~\cite{IllinoisGRMHD:web}.

We choose two different
supercomputers to test the consistency of
\code{Einstein Toolkit} and to obtain an estimate for the sensitivity of
results on
the runtime environment. Consistency in our test is defined as a simulation that uses the same parameter file, and that is created using the same version of the simulation software (\etk) run in different runtime environments, such as different compiler versions, hardware configurations, etc., but generating \textbf{consistent numerical results for physical quantities}. For example, we expect that the central density of the star oscillates with the same amplitude and frequency for simulations on Stampede2 and Expanse, but with slight differences in the numerical results due to compiler optimization, code versions and CPU model.
Thus a bitwise notion of reproducibility is not useful in
this context and instead a relaxed notion of reproducibility based on minimal
expected changes in results due to roundoff errors is used. Roundoff error in this paper is defined the same way as in \ilgrmhdpaper \cite{Etienne:2015cea}. That is, our simulated result should agree with the claimed result at least as well as when simulating otherwise identical data perturbed by the round-off error of the underlying floating point format. More details are discussed in \ref{subsubsec:sig_agreement} and figure \ref{fig:sig_agreement}.

Both Expanse and Stampede2 are supported by
\code{Einstein Toolkit}'s \code{Simulation Factory}~\cite{simfactory:web} module,
which contains information
on how to compile code and submit simulations using the clusters' resource
management system. \code{Simulation Factory} is \code{Einstein Toolkit}'s primary means
to maintain compatibility with computing clusters,
simplifying deployment of code on supported clusters.

\subsection{Experimental setup}

Both \ilgrmhd 2015 and \ilgrmhd 2022 were compiled using
Simulation Factory of \ilgrmhd 2022.
This is required to account for changes in the cluster environment. Both SDSC Expanse and TACC Stampede2 came online after 2015, so none of the compilation instructions of these two clusters is present in \ilgrmhd 2015.

In addition, intermediate versions of \code{IllinoisGRMHD} obtained from the
source code repository were compiled to pinpoint the exact commit that
introduces any significant changes in output.

On all
clusters the code was compiled with value-unsafe optimizations enabled implying
slightly different realizations of each mathematical expression in compiled
code, both between different clusters and between different compiler versions.
The value-unsafe optimizations are achieved by setting various \code{OPTIMISE}
flags in the machine configuration file included in the Simulation Factory.
The default options for Expanse and Stampede2 configuration enabled value-unsafe optimizations,
which is what we used. In addition, the \ilgrmhdpaper used value-unsafe optimization in their simulations,
which we are comparing with.

\begin{table}[htbp]
    \centering
    \begin{tabular}[t]{|c}
    \hline
    \diagbox{Parameters}{Cluster} \\
    \hline
    CPU \\ 
    \hline
    \# of cores per node \\
    \hline
    Compiler \\ 
    \hline
    Optimization \\ 
    \hline
    MPI \\
    \hline\hline
    \# of nodes \\
    \hline
    \# of MPI ranks \\
    \hline
    \# of OpenMP threads \\
    \hline
    \end{tabular}%
    \vline
    \begin{tabular}[t]{c}
    \hline
    Expanse \\
    \\
    \hline
    AMD EPYC 7742 \\ 
    \hline
    128 \\ 
    \hline
    GCC@10.2.0 \\ 
    \hline
    -O2 -mavx2 -mfma \\
    \hline
    openmpi/4.0.4 \\
    \hline\hline
    1 \\
    \hline
    32  \\
    \hline
    4  \\
    \hline
    \end{tabular}%
    \vline
    \begin{tabular}[t]{c|}
    \hline
    Stampede2-skx \\ 
    \\
    \hline
     Intel Xeon Platinum 8160 \\ 
     \hline
     48 \\ 
     \hline
     Intel@18.0.2 \\ 
     \hline
     -Ofast -AVX512 -xHost \\
     \hline 
     impi/18.0.2 \\
    \hline\hline
     1  \\
     \hline
     24  \\
     \hline
     2  \\
    \hline
     
    \end{tabular}
    \caption{ 
    Top half: Hardware, compiler, and optimization level for each cluster. Bottom half: compute resources used in test simulation. On Expanse we used nodes in the ``compute'' partition, and on Stampede2, we used the nodes on ``stampede2-skx'' partition. Nodes on these two clusters have different hardware configurations. MPI library names, and optimization level flags are taken from \code{Simulation Factory} in \ilgrmhd 2022. Differences such as these in hardware, compile and runtime environment may cause observable differences in the output of physical quantities even for identical input.}
    \label{tab:compiler}
\end{table}

Cluster configurations and compiler versions are shown in table \ref{tab:compiler}. Compiling \code{IllinoisGRMHD} on the two clusters is slightly different since compiler versions and queuing systems infrastructure differ between the two clusters.
In each case, we use settings taken from \code{Simulation Factory} in
\ilgrmhd 2022, which supports both clusters. The key difference
between the clusters, for \code{IllinoisGRMHD}, is the different CPU
microarchitecture used: AMD EPYC, launched in 2017, and Intel Skylake, launched
in 2015. This, combined with different compilers and
aggressive optimization settings used, results in round-off level differences
when evaluating mathematical expressions. These differences then propagate
and, potentially, could amplify to levels incompatible with consistent physical results. On
the other hand, different MPI stacks on clusters, the MPICH-based Intel MPI
stack on Stampede2 and OpenMPI on Expanse, do not influence the numerical results since
\code{IllinoisGRMHD}'s evolution code solely uses data transfer primitives,
e.g., \code{MPI\_Send} and \code{MPI\_Recv}, that only copy data identically
but not on reduction operations that act on values.

The code published in~\cite{Etienne:2015cea} does not include scripts to
post-process the raw simulation output and plot graphs shown in the manuscript.
As part of the experiment, the required scripts were implemented in Python
based on information available in the published material. Additionally, copies
of the original scripts were obtained from the
author and are now available without modification on \cite{zachariah_etienne_2022_7545717}. Their output, given identical input files, was
compared to that of the newly implemented Python code.

\subsection{Simulation parameters and diagnostics}

\etk simulations are controlled via parameter files, which define numerical
simulation inputs, such as grid spacing, evolution method, and
initial
data of the physics setup. \etk targets backward compatibility of parameter files -- a parameter file
run using Einstein Toolkit
version \ilgrmhd 2015 should produce numerically consistent results to that same parameter file run using any later version such as \ilgrmhd 2022.

We reproduce one of the tests
in \ilgrmhdpaper using the original parameter file to verify this. Two simulations were
created: one using \ilgrmhd 2015 and the other using
\ilgrmhd 2022. The parameter file, \code{tov\_star\_parfile\_for\_IllinoisGRMHD.par}, used as the basis for this experiment, is included in \ilgrmhd 2015 and was used with only modifications on the grid spacing corresponding to different resolutions.

All simulations use a cubic fixed mesh refinement grid, and x, y, z dimensions all have the same
number of grid points. The grid spacing in high, medium, and low-resolution
simulations are $(0.32, 0.4, 0.5)$ code units, respectively, in the coarsest refinement level. Four
refinement levels of size $(1.5, 3.0, 6.0, 12.0)\ R_{NS}$, where $R_{NS}$ is the radius of the star, are used. Therefore, the grid spacings in the finest refinement level are
$\Delta x_{finest}=(0.02, 0.025, 0.03125)$ code units, which correspond to (75, 60, 48) grid points inside the finest refinement level.

We follow the boundary conditions as defined in \cite{Etienne:2015cea}.
For primitive variables ($\rho_0$, $P$, $v^{i}$), the outer boundary enforces zero-derivative, value-copy of these primitive variables to the outer boundary if the incoming velocity is positive,
which is referred to as ``outflow'' boundary conditions in \cite{Etienne:2015cea}.
The outer boundary condition for vector potentials $A_\mu$ and $\sqrt{\gamma}\Phi$ is to linearly extrapolate the values to the outer boundary.

Following the conventions in \cite{Etienne:2015cea}, we use two outputs, namely, the change in central density ($\Delta \rho_c$) and the L2-norm of the
Hamiltonian constraint violation ($\|\mathcal{H}\|$) in the numerical tests.
Change in central density is defined as
\begin{equation}
\Delta \rho_c = \frac{\rho_c(t) - \rho_c(t=0)}{\rho_c(t=0)} \,\text{.}
\end{equation}
For a stable TOV star in equilibrium, both the change in central density and
the Hamiltonian constraint violation are expected to converge to zero when resolution increases. For finite resolution both $\Delta \rho_c$ and $\|\mathcal{H}\|$ will have a small but nonzero value.

Convergence order
is used in~\cite{Etienne:2015cea} to evaluate the performance of the code for
different resolutions and as a basic check on the correct implementation of the
evolution equations.
For a quantity $Q \in \{\Delta \rho_c, \|\mathcal{H}\|\}$ whose expected value is $0$,
convergence order for a set of
two resolutions $\Delta x_1$ and $\Delta x_2$ is computed as  
\begin{equation}
    n = \log\left(\frac{Q(\Delta x_1)}{Q(\Delta x_2)}\right) / \log\left(\frac{\Delta x_1}{\Delta x_2}\right) \mbox{,}
\end{equation}
where $Q(\Delta x_{1,2})$ is the quantity as computed in the simulation with
grid spacing $\Delta x_1$ and $\Delta x_2$, respectively.

We use two additional quantities to measure the difference in numerical results between different setups. The absolute difference in change of central density is defined as
\begin{equation}
    \Delta^{\text{abs}}(\Delta \rho_{c}) = |\Delta \rho_{c,1} - \Delta \rho_{c,2}|
\end{equation}
and the relative difference in Hamiltonian constraint violation is defined as
\begin{equation}
    \Delta^{\text{rel}}(\|\mathcal{H}\|) = \frac{\bigg| \|\mathcal{H}\|_{1} - \|\mathcal{H}\|_{2}\bigg|}{\|\mathcal{H}\|_{2}}
    \,\text{,}
\end{equation}
where relative differences are used for $\|\mathcal{H}\|$ to remove dependencies on an
arbitrary overall scale for $\|\mathcal{H}\|$.
In order to perform this comparison between simulations, one of the time series in the comparison may have to be interpolated to match the time steps of the other time series.
Time series $1$ always has a higher sampling rate than time series $2$, and we linearly interpolate time series $1$ to match the timesteps of time series $2$. Because both time series have very high sampling rates and we downsample time series $1$ to match the other, the comparison results should not be affected by the interpolation qualitatively.
Additional details are specified in the caption of each figure.

\subsection{Results}\label{sec:results}

After simulating the test cases included in the \ilgrmhd 2015
announcement manuscript~\cite{Etienne:2015cea}, we compared our simulation results
with those results in~\cite{Etienne:2015cea}.  In particular, the Hamiltonian constraint
$\|\mathcal{H}\|$, change in central density $\Delta \rho_c$, and convergence order of
these quantities are used to compare results across
different supercomputers and to the original publication results in figure 3
of~\cite{Etienne:2015cea}. 

This section presents our reproducibility study results by first computing results using two different versions of the \etk running the same parameter file once on each of TACC Stampede2 and SDSC Expanse, respectively. Then the simulation results by the same version of the \etk on two different clusters are compared against each other, as well as different \etk versions on the same cluster.

The simulations on both clusters were run until at least $t = 55\, t_{dyn} \approx 153.0$, where  $t_{dyn}=\sqrt{1/\rho_c}$, and $\rho_c \approx 1.29\times10^{-1}$ is the central density of the star at $t=0$ in $G=c=M_\odot=1$ unit.
The numerical simulation setup and running steps of this reproducibility experiment are reproducible with details in \ref{sec:compilecode}. Our simulation setup files, raw data, and analysis code are available at \wyodoi.

\subsubsection{Round-off level agreement between simulations}\label{subsubsec:sig_agreement}

\begin{figure}
    \centering
    \includegraphics[width=.9\linewidth]{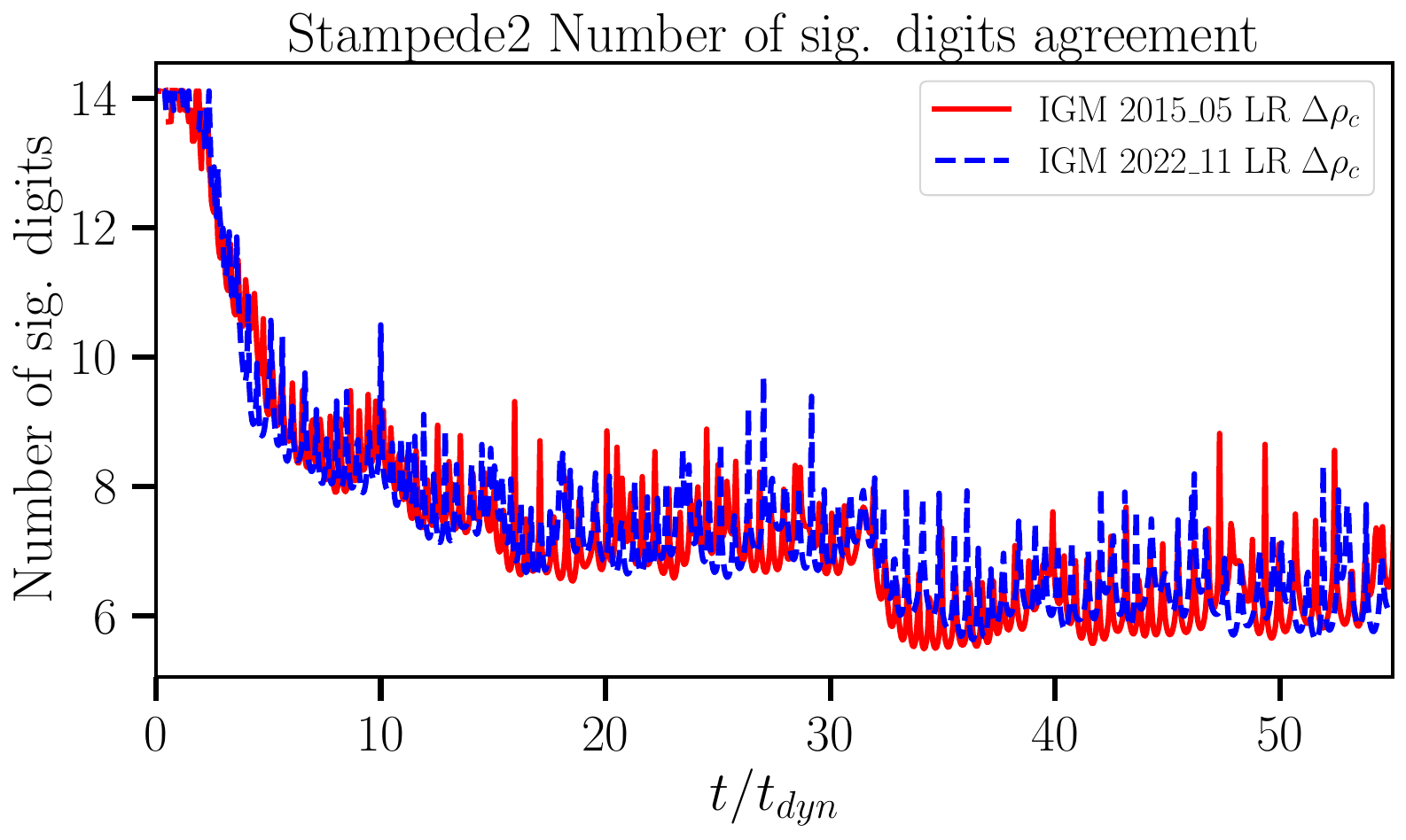}
    \caption{The number of significant digits agreement for central density changes between initial 15\textsuperscript{th} digit perturbed and unperturbed results. The red solid line is the significant digits agreement between initially perturbed results with \ilgrmhd2015 and \ilgrmhdresult. The Blue dashed line is the comparison between initially perturbed and unperturbed results with \ilgrmhd2022. Our \ilgrmhd results are obtained on Stampede2.
    }
    \label{fig:sig_agreement}
\end{figure}

In~\cite{Etienne:2015cea} a notion of agreement up to round-off
errors is introduced by observing how two simulations, using the same executable
code, whose input parameters relative difference is no more than the
floating point $\epsilon$ (see \ref{sec:roundoffdifferences} for details)
deviate from each
other over the course of the simulation. To understand the round-off error of our numerical experiments, we performed the same significant digits agreement test as in \cite{Etienne:2015cea}.

An initial 15\textsuperscript{th} digit random perturbation was added to the initial data on the grid in manner described in \ilgrmhdpaper and repeated in appendix~\ref{sec:roundoffdifferences}. That is, at each grid point all primitive variables in IllinoisGRMHD are multiplied by a common factor $1+\epsilon$, where $\epsilon$ is a random number in the interval $[0, 10^{-15})$. All conserved physical quantities are re-calculated based on the new perturbed initial primitive variables. After 30 dynamical timescales, the significant digits agreement for both \ilgrmhd2015 and \ilgrmhd2022 cases oscillate between 6 -- 8 digits. Our results, as shown in figure \ref{fig:sig_agreement}, agree with the original publication result in \cite{Etienne:2015cea} figure 1. Hence, for the set of tests considered in this manuscript, two simulations that differ by no more than an absolute error of $10^{-6}$ in $\rho$ after 30 dynamical time scales are considered to agree up to round-off level errors.

\subsubsection{Expanse} \label{sec:result_expanse}

\begin{figure}
    \begin{subfigure}{.55\textwidth}
      \centering
      \includegraphics[width=.95\linewidth]{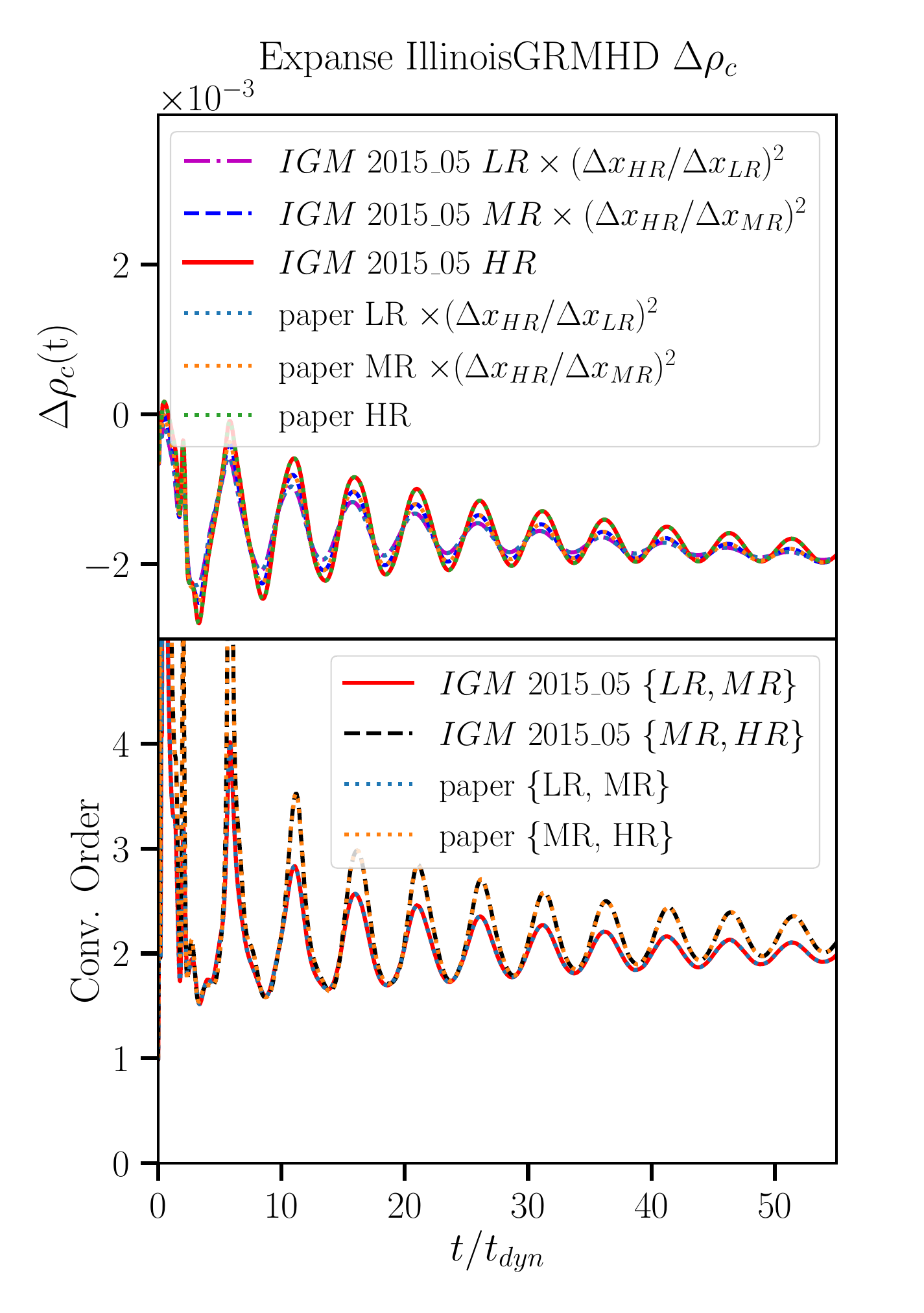}
      \label{fig:expanse2015rho}
    \end{subfigure}%
    \begin{subfigure}{.55\textwidth}
      \centering
      \includegraphics[width=.95\linewidth]{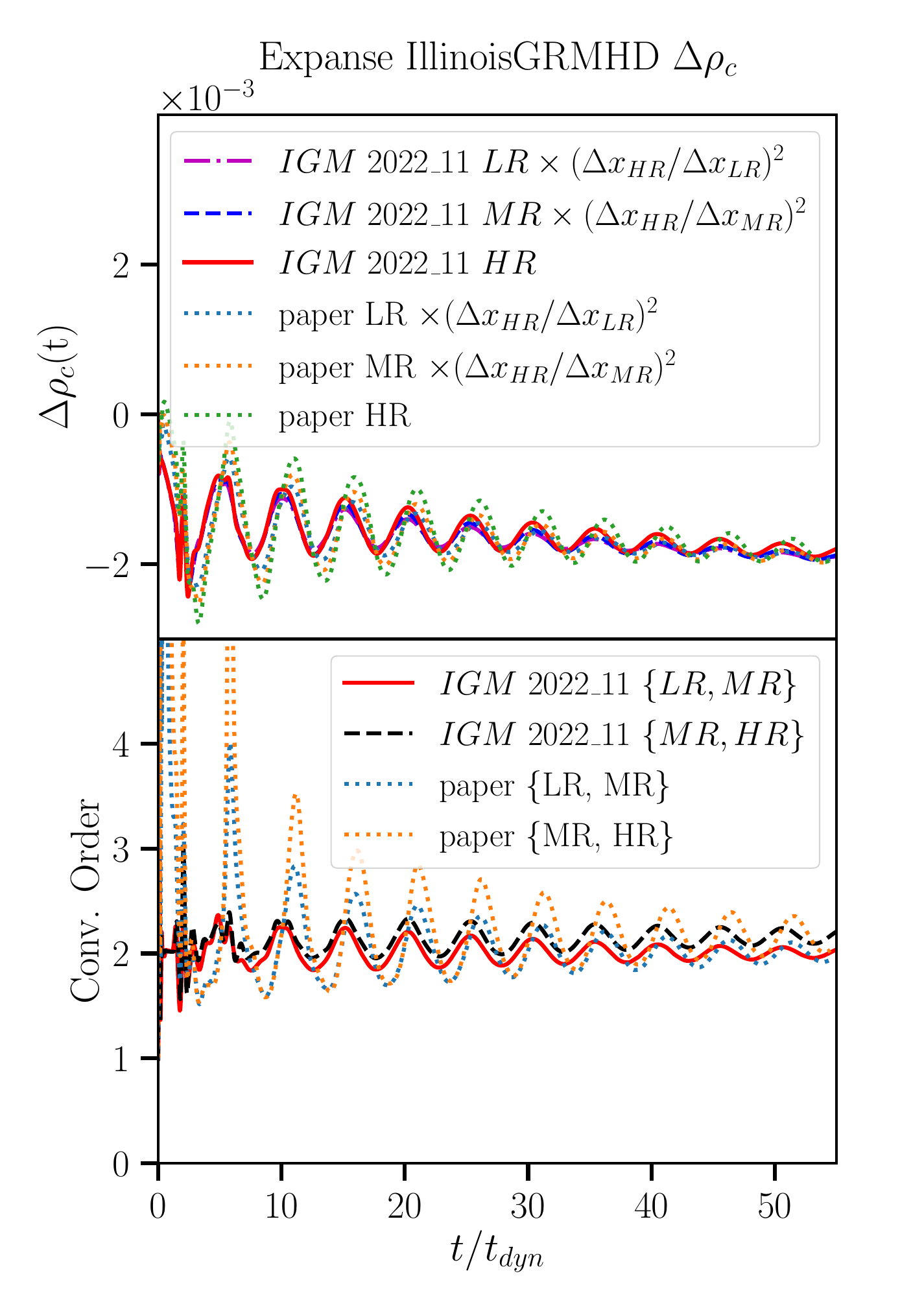}
      \label{fig:expanse2022rho}
    \end{subfigure}%
    \\
    \begin{subfigure}{.55\textwidth}
      \centering
      \includegraphics[width=.95\linewidth]{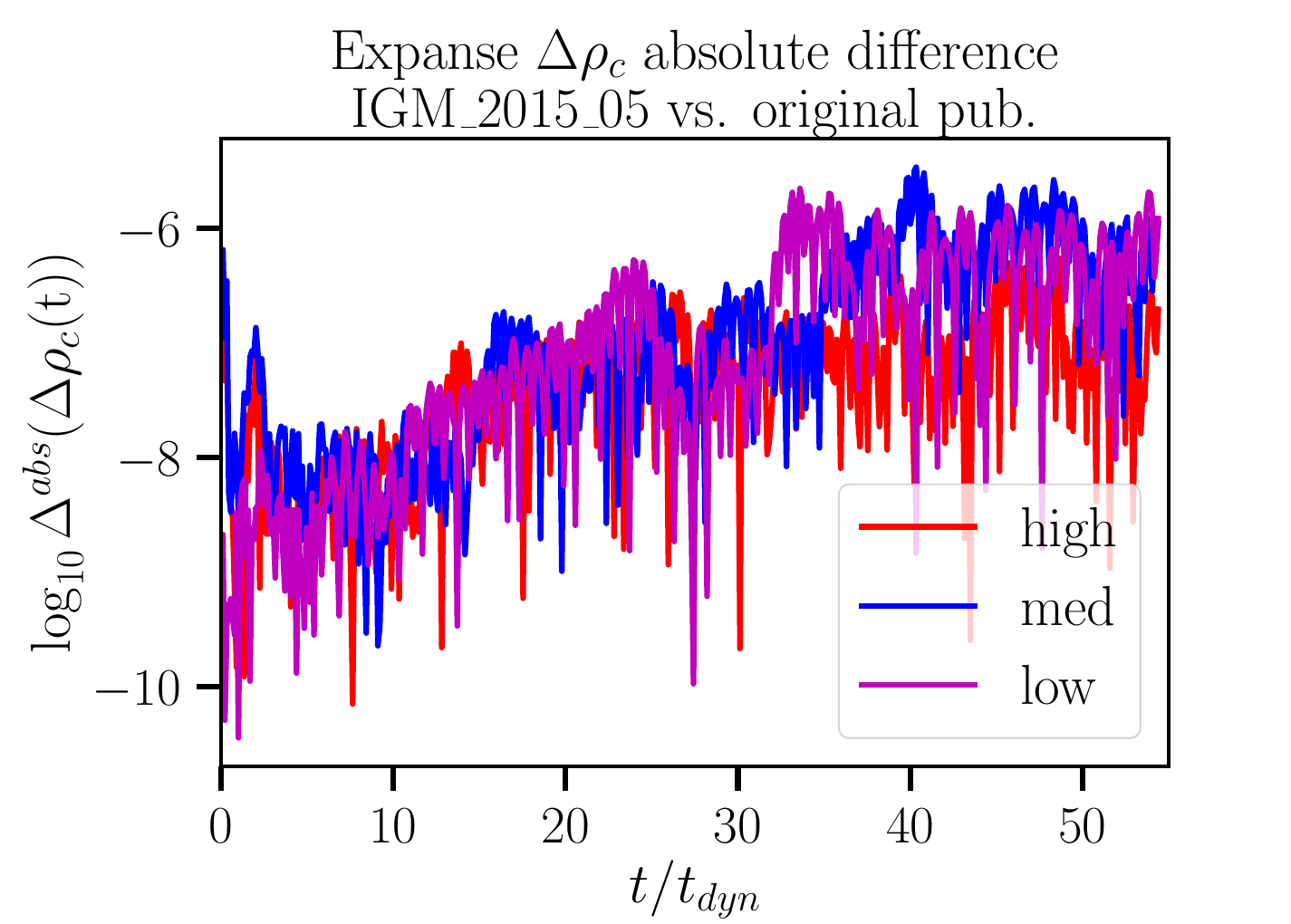}
      \caption{\ilgrmhd 2015}
      \label{fig:expanse2015rhodiff}
    \end{subfigure}%
    \begin{subfigure}{.55\textwidth}
      \centering
      \includegraphics[width=.95\linewidth]{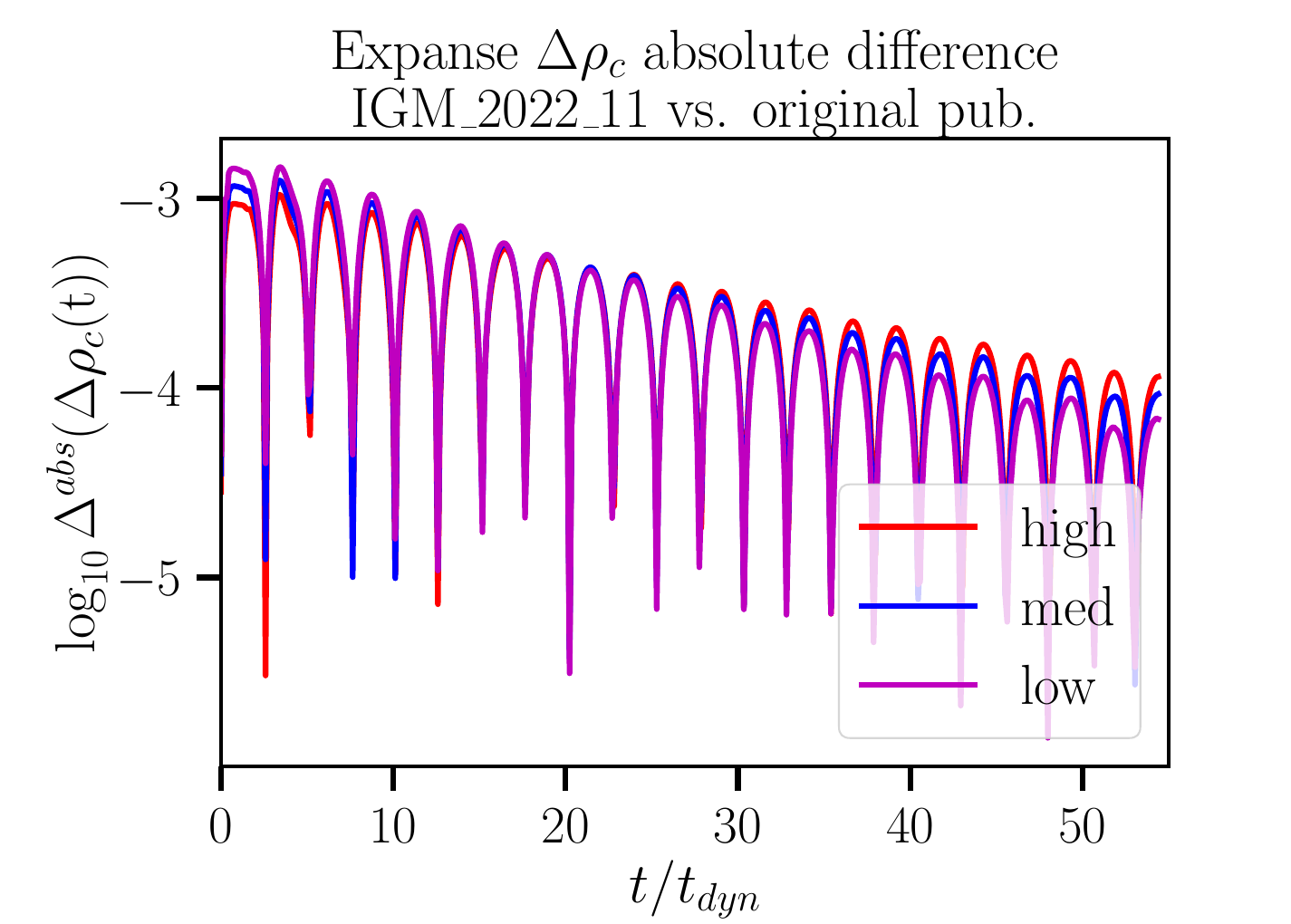}
      \caption{\ilgrmhd2022}
      \label{fig:expanse2022rhodiff}
    \end{subfigure}%
    \caption{Comparing the central density changes $\Delta \rho_c$ for simulations created with two versions of the \etk on Expanse. The left two panels are for the \ilgrmhd 2015, while the right two are for the \ilgrmhd 2022. The upper two figures show $\Delta \rho_c$ and its convergence order. The bottom figures present the magnitude of the absolute error between our results and \ilgrmhdresult. The right column figures show there is a significant difference between \ilgrmhd2022 result and \ilgrmhdresult.}
    \label{fig:expanse_rho}
\end{figure}

Figure \ref{fig:expanse_rho} shows the agreement between our simulation result for $\Delta \rho_c$ and the original \ilgrmhd published result. 
The top row shows how $\Delta \rho_c$ evolves when simulated using \ilgrmhd2015
and \ilgrmhd2022 and compares to the results of \ilgrmhdpaper[]. On the left, for
\ilgrmhd2015, which uses the code published in \ilgrmhdpaper[], our results, shown
in solid, dashed, and dot-dashed lines, stop tracking the (dotted) results of 
\ilgrmhdpaper[]. This is especially evident in the convergence order plot in the
middle section, which shows nearly perfect agreement. This is shown explicitly in
the bottom panel, which demonstrates that our results and \ilgrmhdpaper differ
by no more than $10^{-6}$ and are consistent within round-off error as 
introduced in section~\ref{sec:roundoffdifferences}.

The right-hand column displays corresponding results comparing \ilgrmhd2022
with \ilgrmhdpaper[]. Here, differences are much more obvious, already in the top
plot for $\Delta \rho_c$, where there is much less overlap visible between the 
curves. This difference is much more obvious in the convergence plot in the 
middle, which shows \ilgrmhd2022 having much smaller oscillations around the 
expected convergence of $2$. The bottom right absolute difference graph 
quantifies this and shows that the difference starts out very large of order
$10^{-3}$ and slowly decreases to $10^{-5}$ as the simulation progresses. 
However, even at the end of the simulation, the difference still exceeds the 
threshold magnitude established for round-off level agreement. This significant
difference can be tracked down to git commit 
\href{https://bitbucket.org/zach_etienne/wvuthorns/commits/8b562af09a888b2d795506e5711cc42a72f840c4}{8b562af09}, which is present in \ilgrmhd2022 but not in \ilgrmhd2015. We have verified
that reverting this single commit brings \ilgrmhd2022 into round-off level 
agreement with \ilgrmhdpaper[]. Inspecting the commit message reveals that the 
commit fixes a minor bug present in \ilgrmhd2015 which results in an incorrect
energy density being present at $t=0$, physically corresponding to an out-of-equilibrium configuration, which explains why the observed difference decreases
over time as the system relaxes back to the equilibrium configuration. We refer to this discrepancy as the incorrect initial stress-energy tensor issue.

\begin{figure}
    \begin{subfigure}{.55 \textwidth}
      \centering
      \includegraphics[width=.95\linewidth]{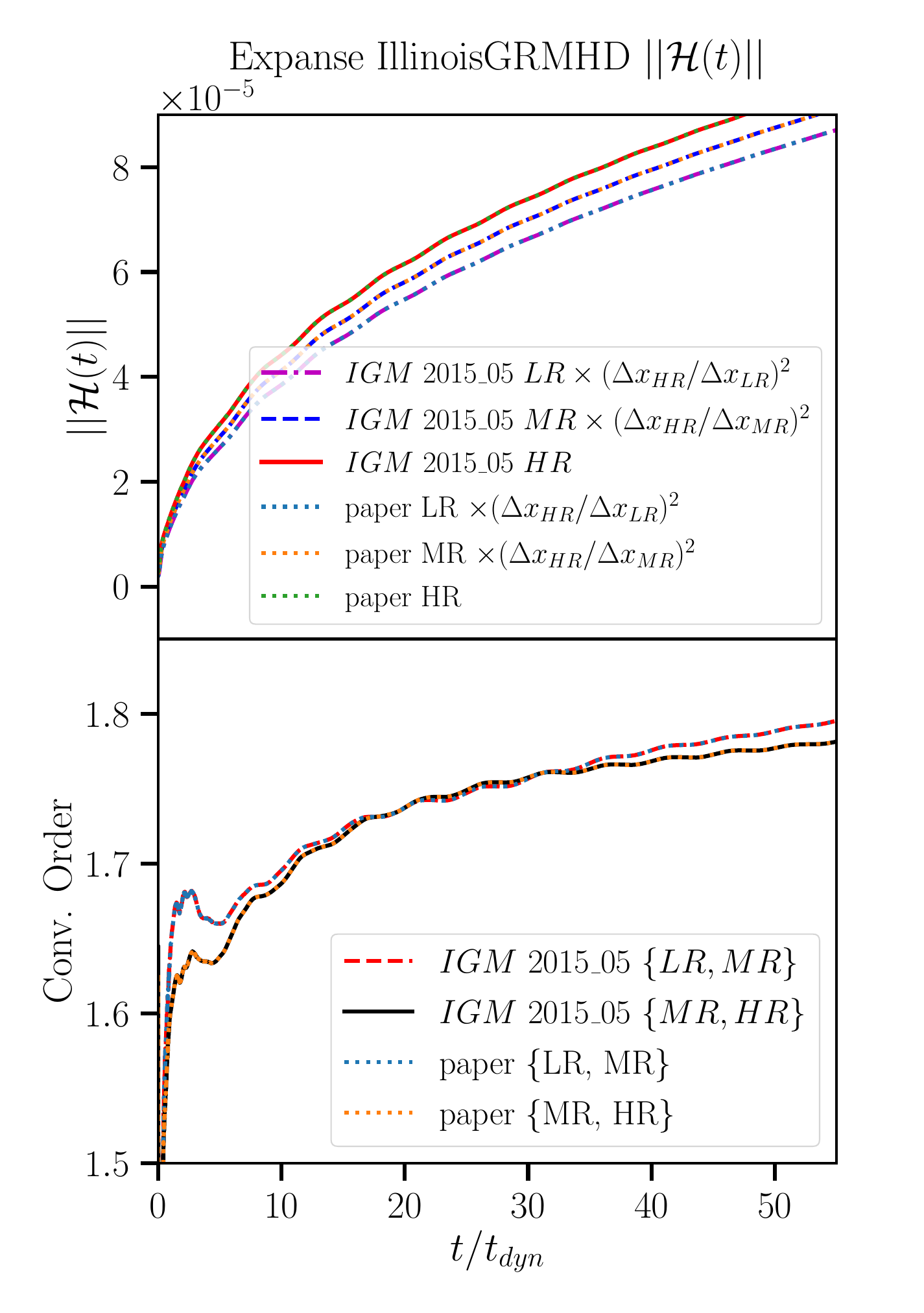}
      \label{fig:expanse2015ham}
    \end{subfigure}%
    \begin{subfigure}{.55\textwidth}
      \centering
      \includegraphics[width=.95\linewidth]{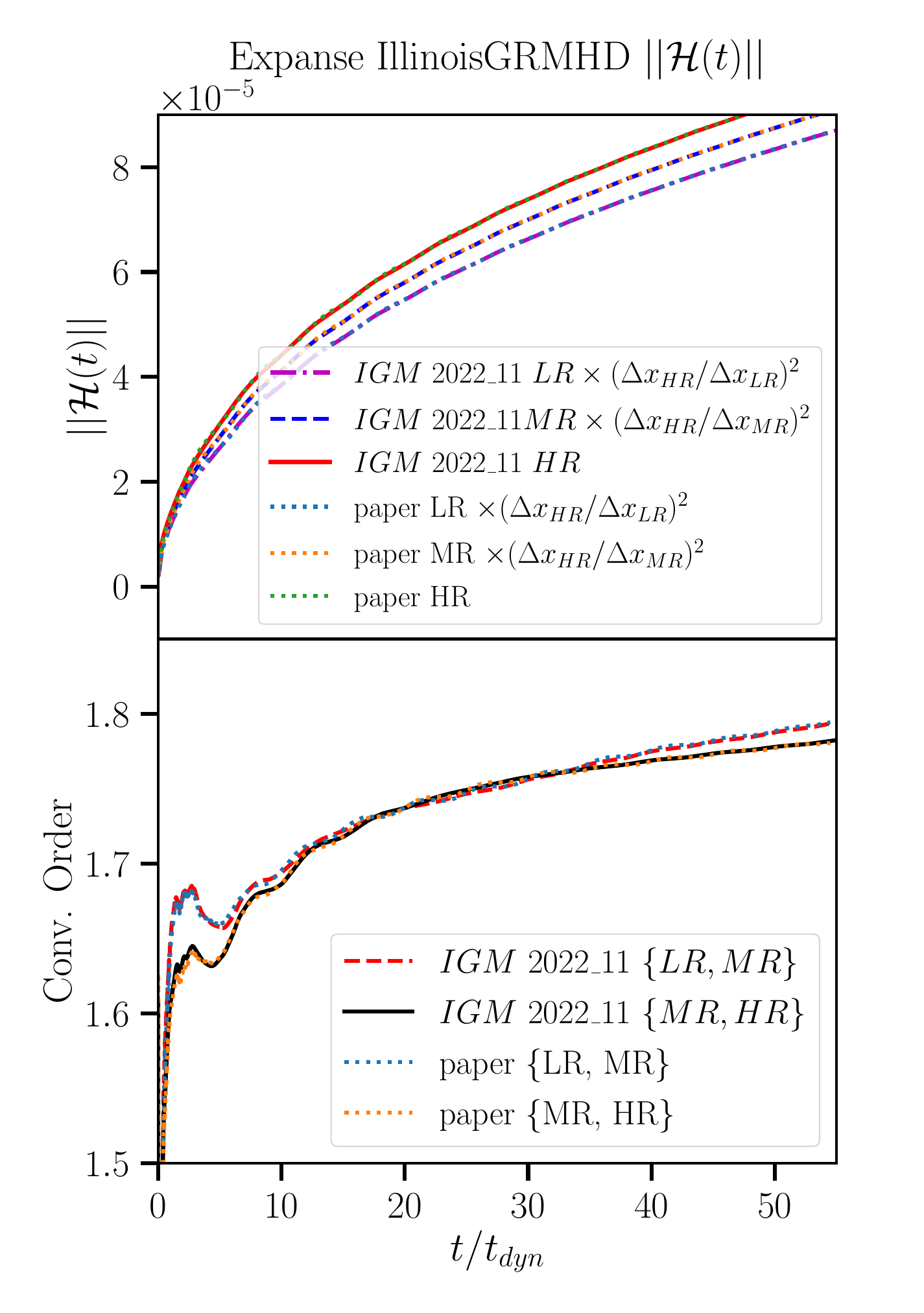}
      \label{fig:expanse2022ham}
    \end{subfigure}%
    \\
    \begin{subfigure}{.55\textwidth}
      \centering
      \includegraphics[width=.95\linewidth]{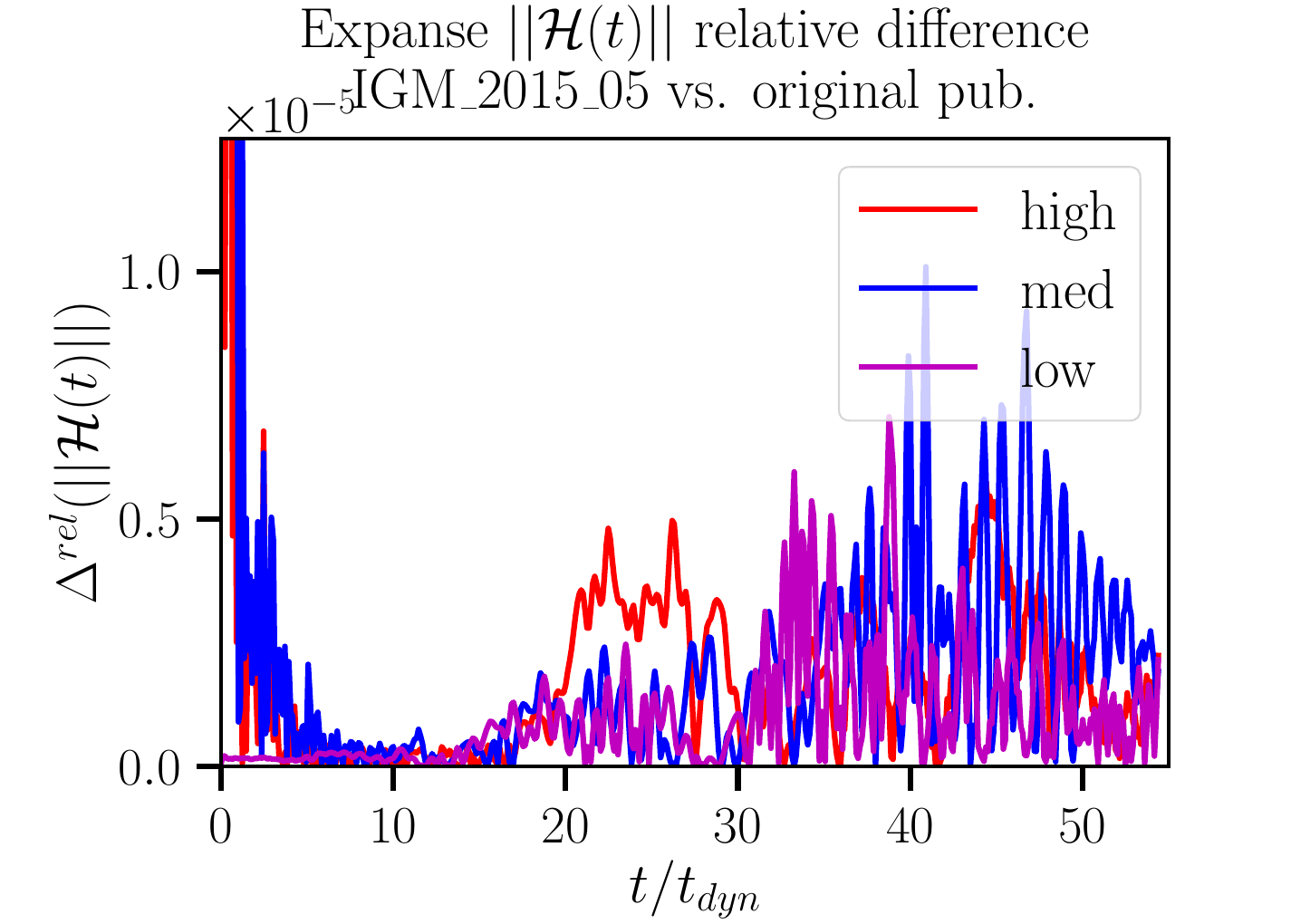}
      \caption{\ilgrmhd 2015}
      \label{fig:expanse2015hamdiff}
    \end{subfigure}%
    \begin{subfigure}{.55\textwidth}
      \centering
      \includegraphics[width=.95\linewidth]{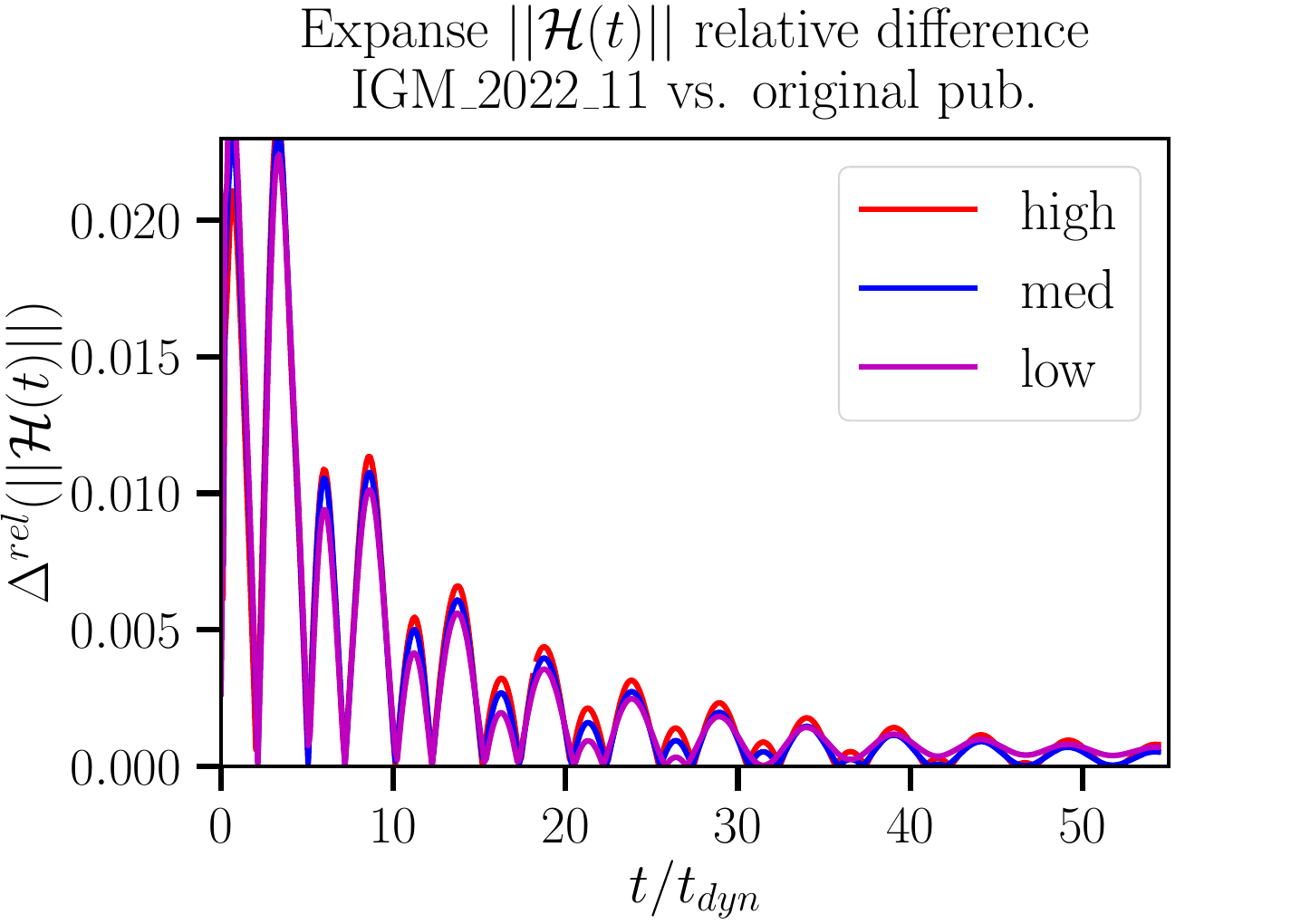}
      \caption{\ilgrmhd 2022}
      \label{fig:expanse2022hamdiff}
    \end{subfigure}%
    \caption{Comparing the Hamiltonian constraint $\|\mathcal{H}\|$ for simulations created with two versions of \etk on Expanse. The left panel is for the \ilgrmhd 2015, and the right panel is for the \ilgrmhd 2022. The upper two figures show $\|\mathcal{H}\|$ and the convergence order. The bottom two present the relative error between our results and \ilgrmhdresult.}
    \label{fig:expanse_ham}
\end{figure}

These same observations also hold in figure~\ref{fig:expanse_ham}
which displays results for the Hamiltonian constraint $\|\mathcal{H}\|$.
Similar to the situation for $\Delta \rho_c$, our result agrees with the originally published result with \ilgrmhd 2015 and has a significant difference for \ilgrmhd 2022, due to the incorrect initial stress-energy tensor issue described in the previous paragraph.

\subsubsection{Stampede2}

\begin{figure}
    \begin{subfigure}{.55\textwidth}
      \includegraphics[width=.95\linewidth]{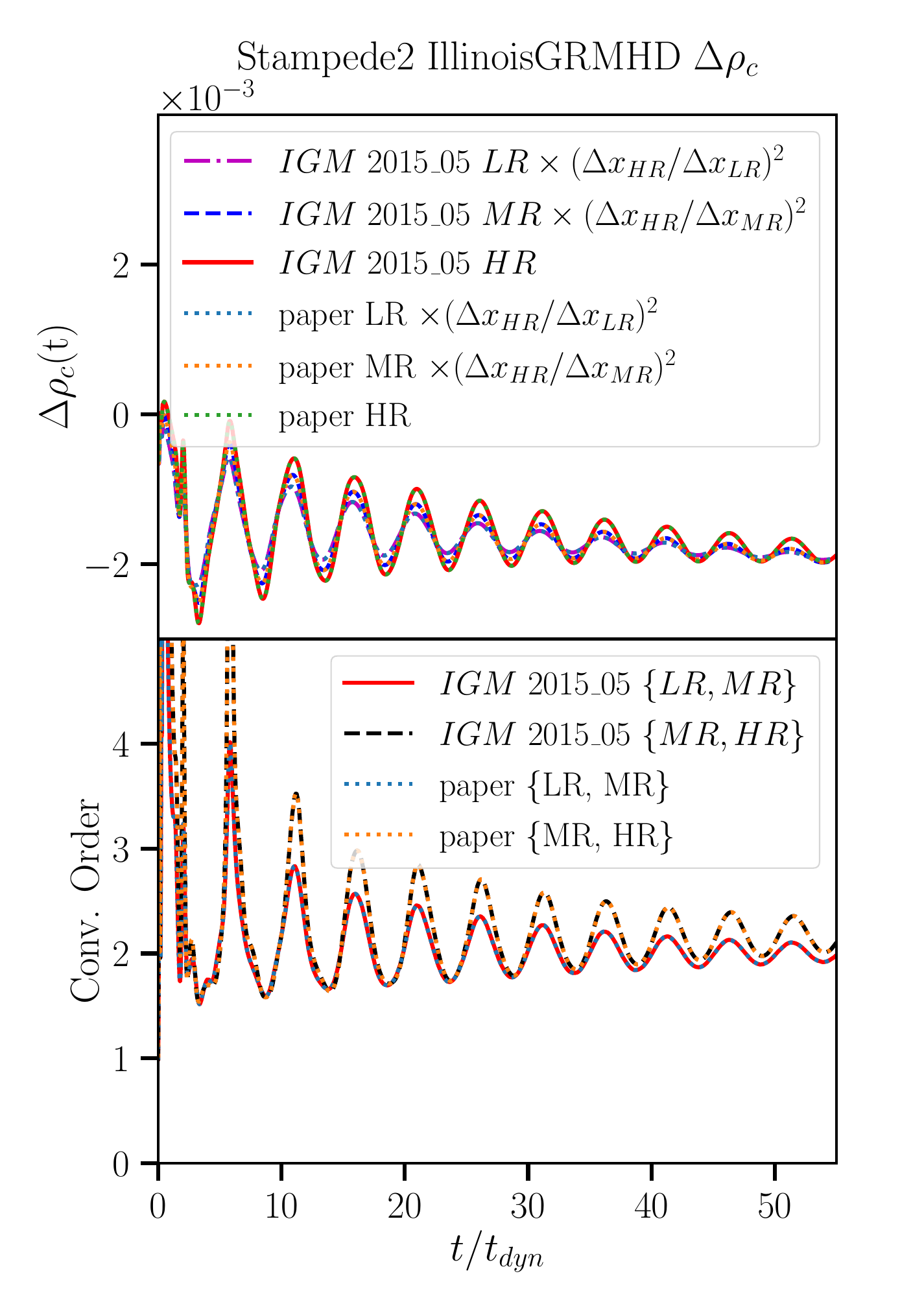}
      \label{fig:sfig1}
    \end{subfigure}%
    \begin{subfigure}{.55\textwidth}
      \centering
      \includegraphics[width=.95\linewidth]{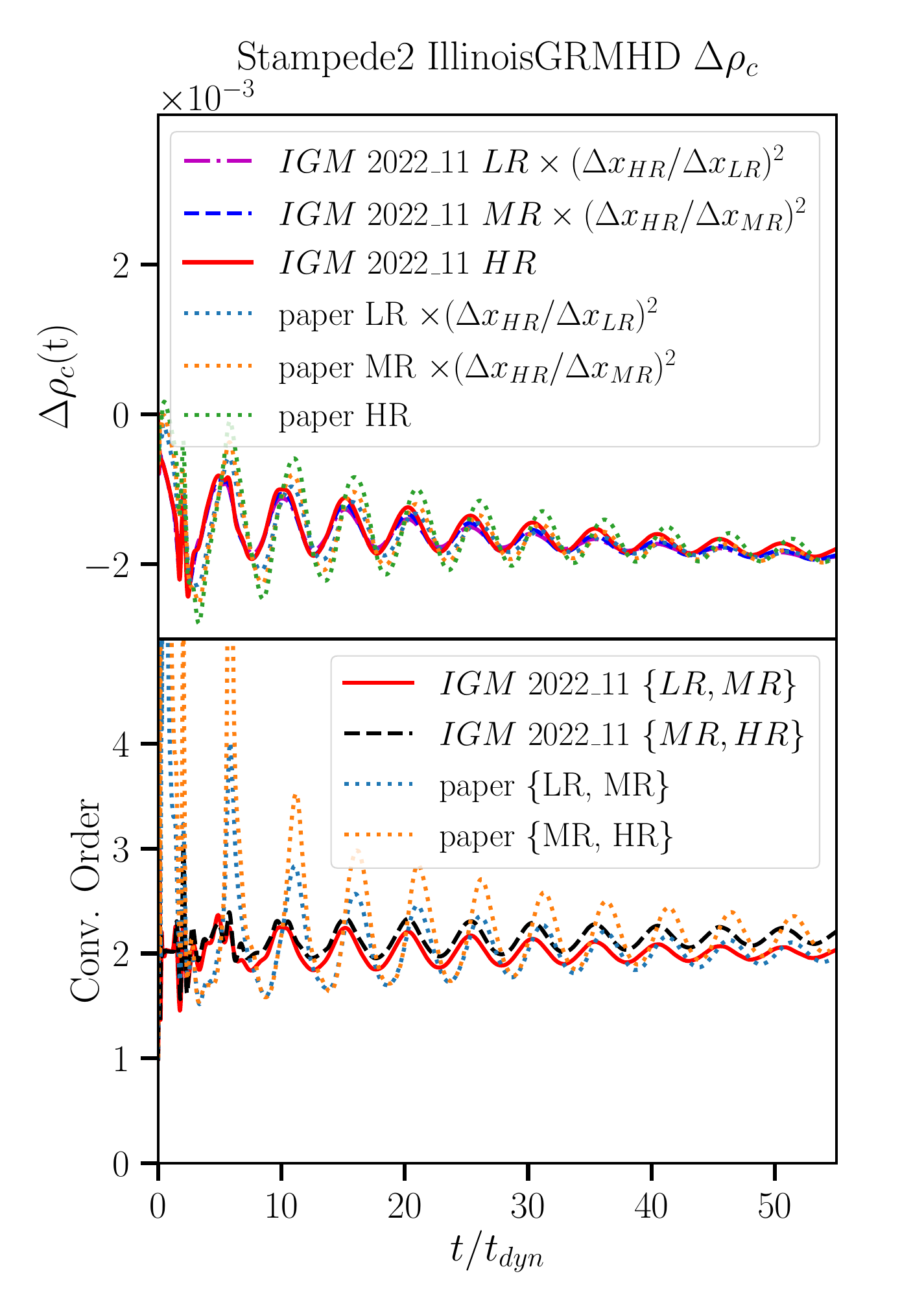}
      \label{fig:sfig1}
    \end{subfigure}%
    \\
    \begin{subfigure}{.55\textwidth}
      \centering
      \includegraphics[width=.95\linewidth]{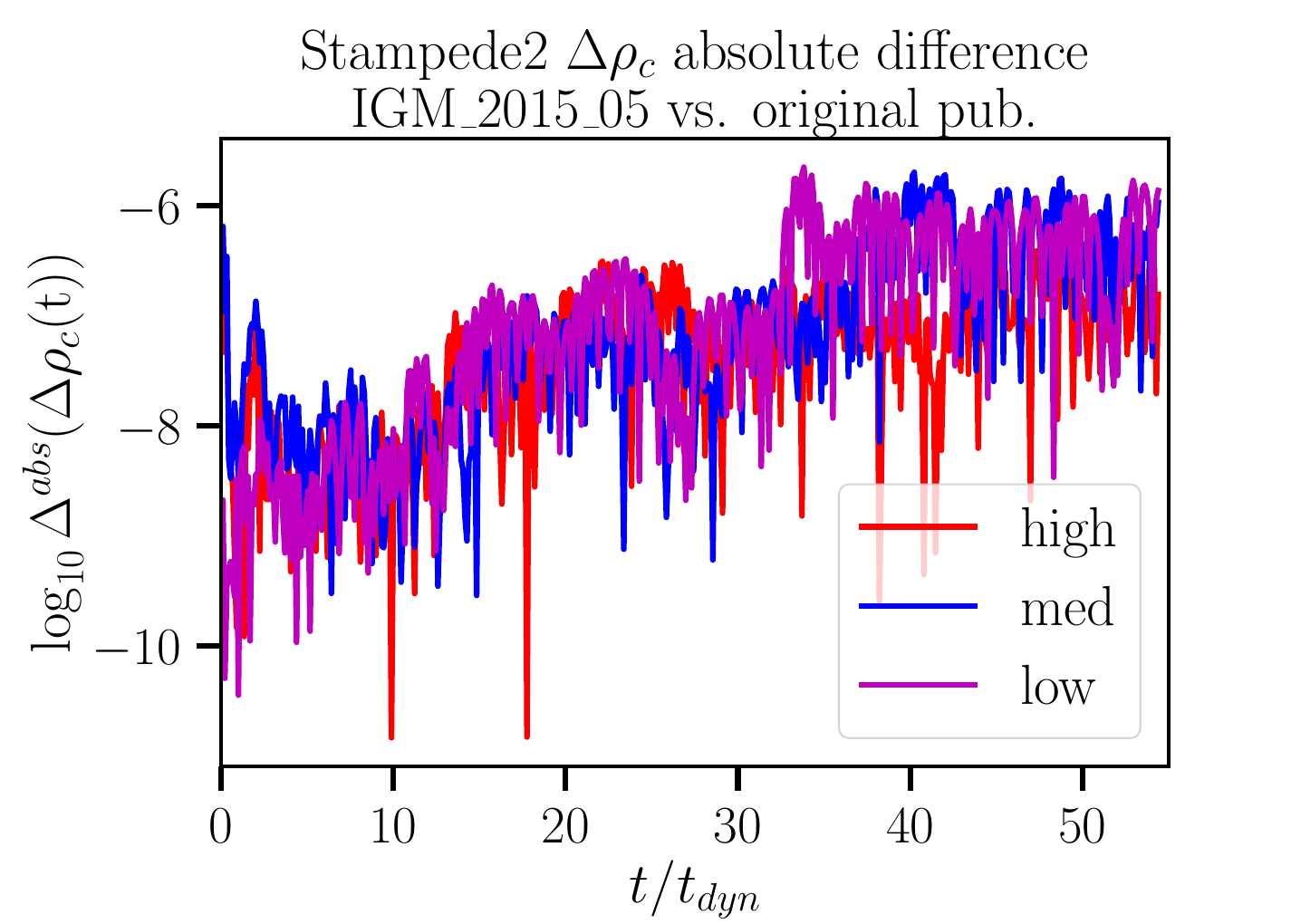}
      \caption{\ilgrmhd 2015}
      \label{fig:sfig1}
    \end{subfigure}%
    \begin{subfigure}{.55\textwidth}
      \centering
      \includegraphics[width=.95\linewidth]{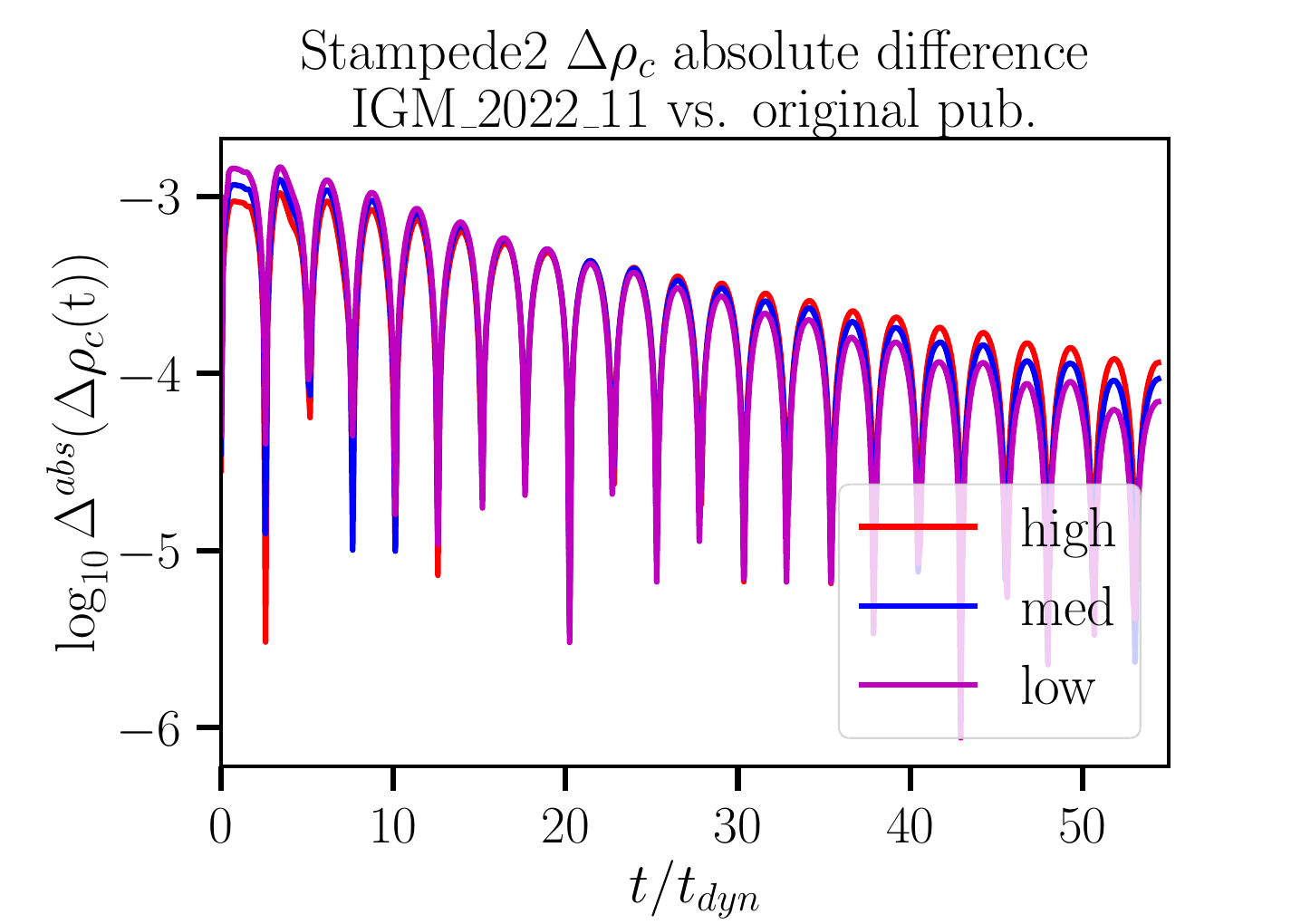}
      \caption{\ilgrmhd 2022}
      \label{fig:sfig1}
    \end{subfigure}%
    \caption{Comparing the central density changes $\Delta \rho_c$ for simulations created with two versions of \etk on Stampede2. This is the same comparison as in figure \ref{fig:expanse_rho}, but instead, the simulation was performed on Stampede2. The significant difference on the right panel is due to the incorrect stress-energy tensor issue described in section \ref{sec:result_expanse}.}
    \label{fig:stampede2rho}
\end{figure}

\begin{figure}
    \begin{subfigure}{.55\textwidth}
      \centering
      \includegraphics[width=.95\linewidth]{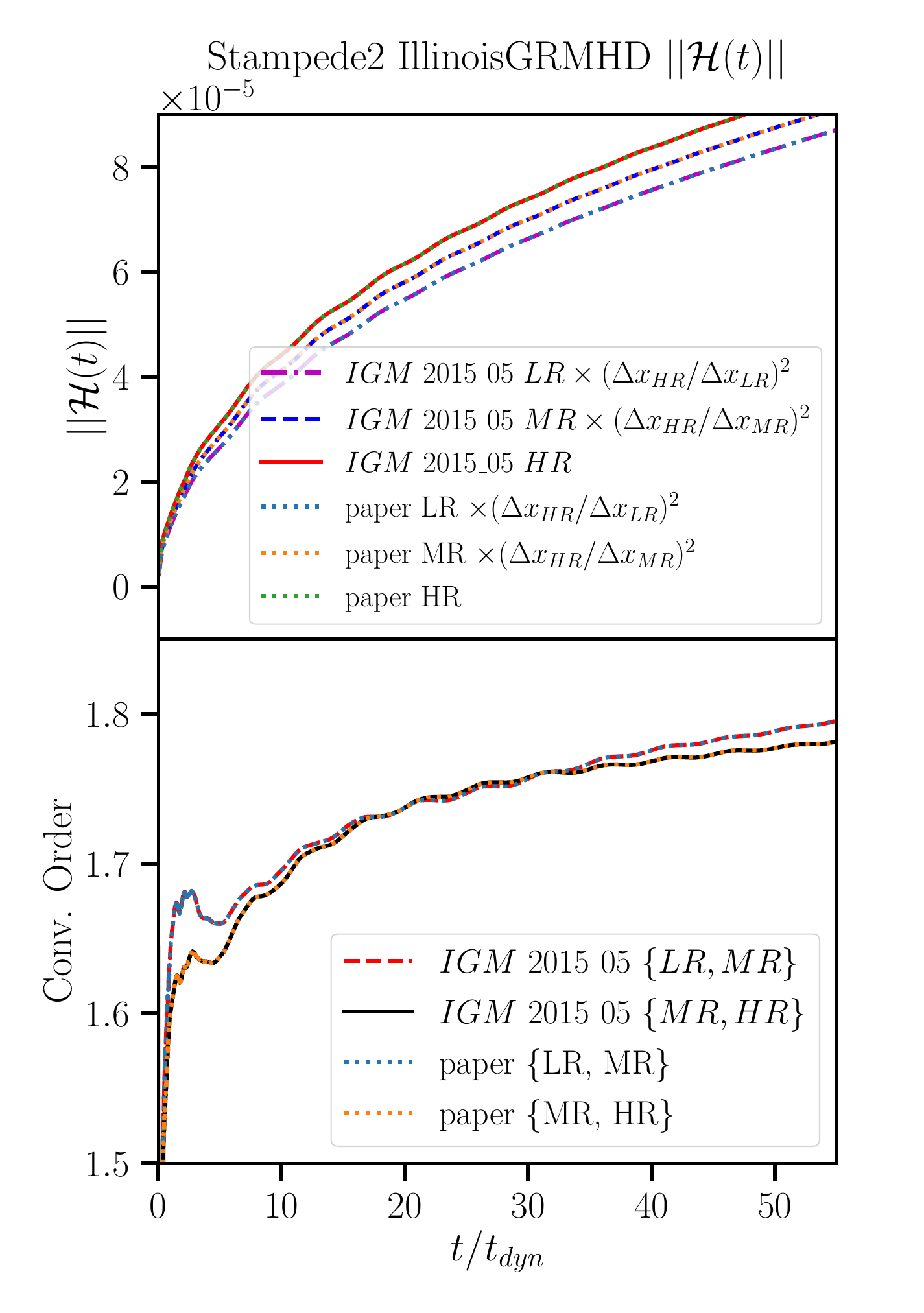}
    \end{subfigure}%
    \begin{subfigure}{.55\textwidth}
      \centering
      \includegraphics[width=.95\linewidth]{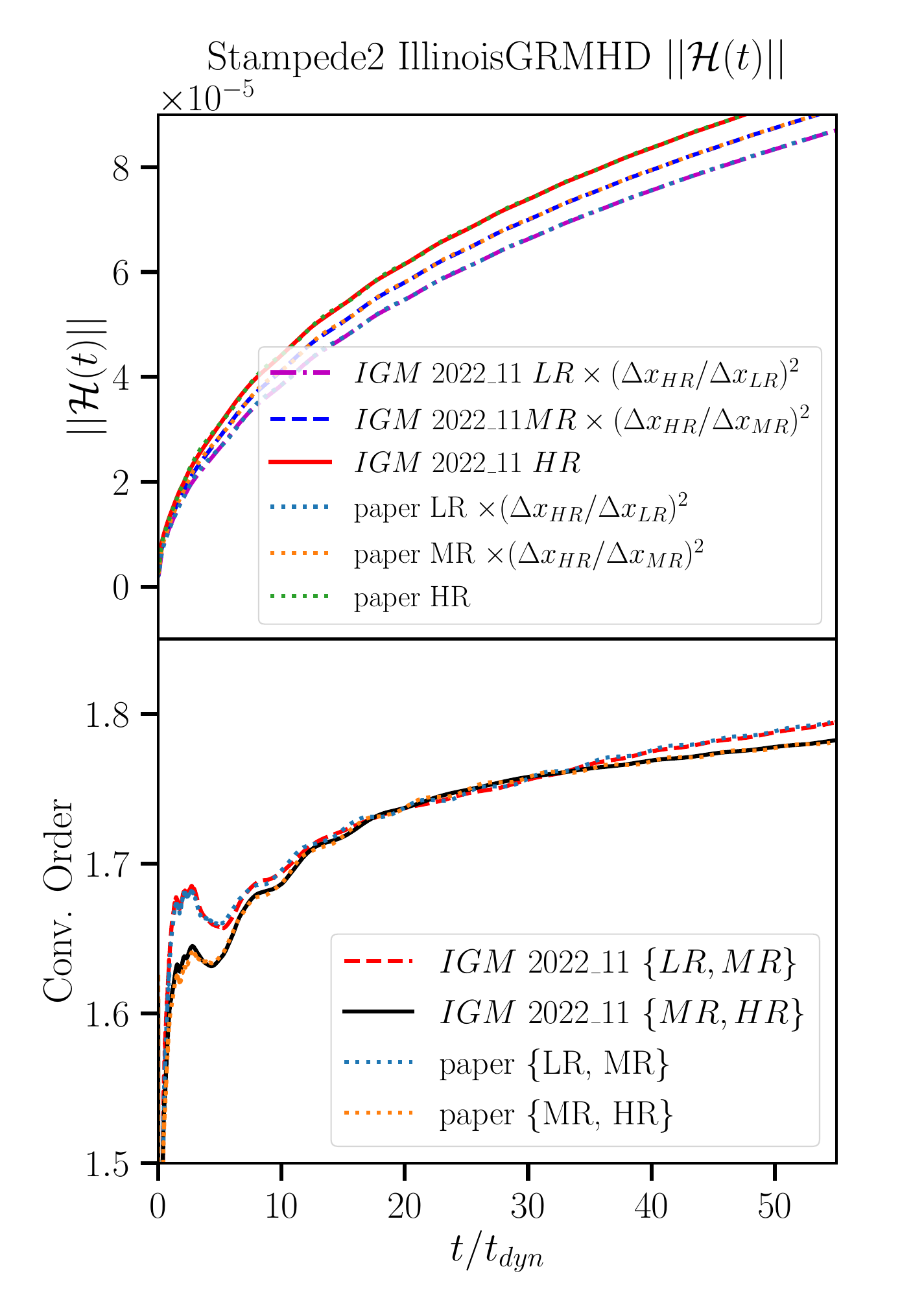}
    \end{subfigure}%
    \\
    \begin{subfigure}{.55\textwidth}
      \centering
      \includegraphics[width=.95\linewidth]{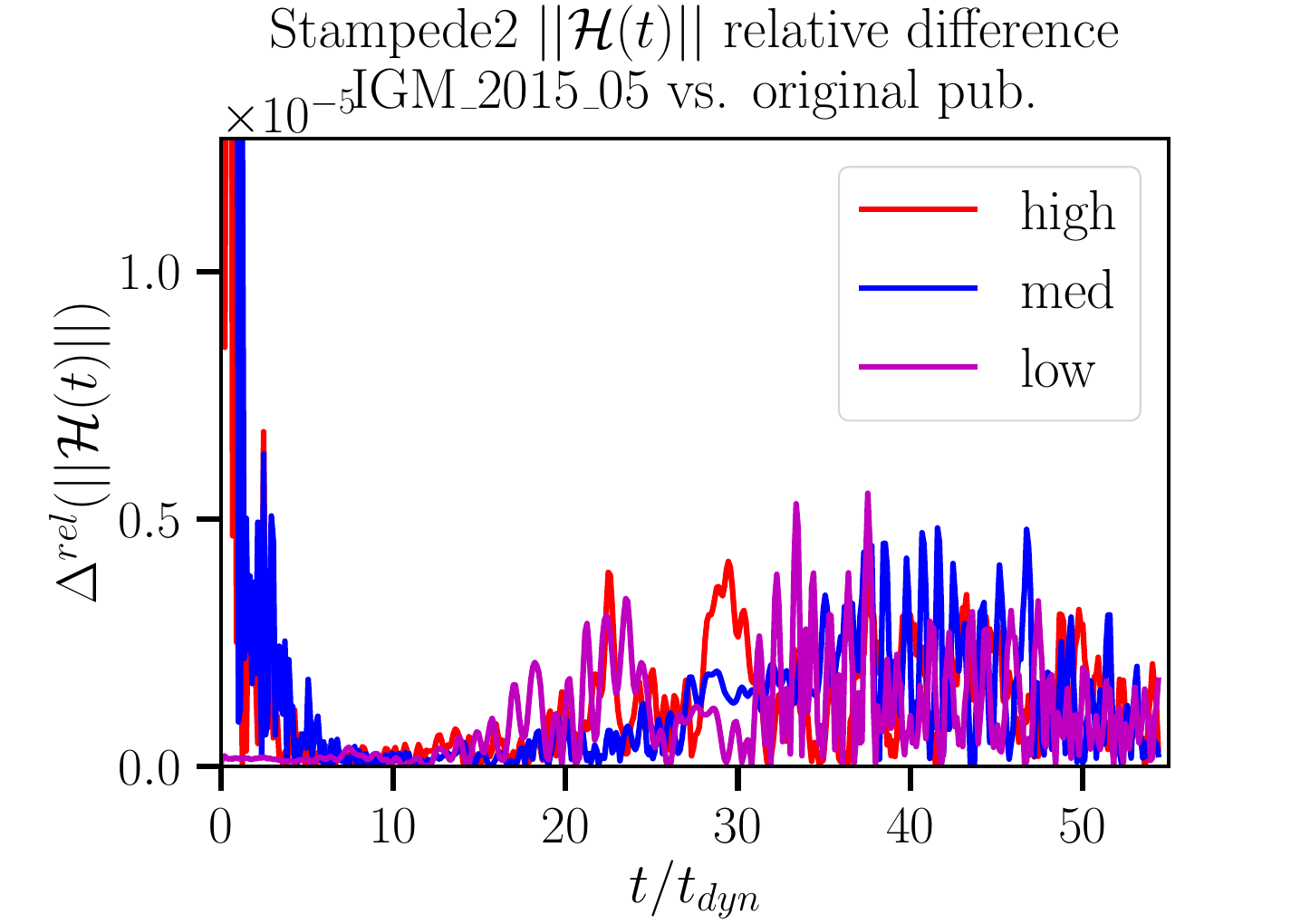}
      \caption{\ilgrmhd 2015}
    \end{subfigure}%
    \begin{subfigure}{.55\textwidth}
      \centering
      \includegraphics[width=.95\linewidth]{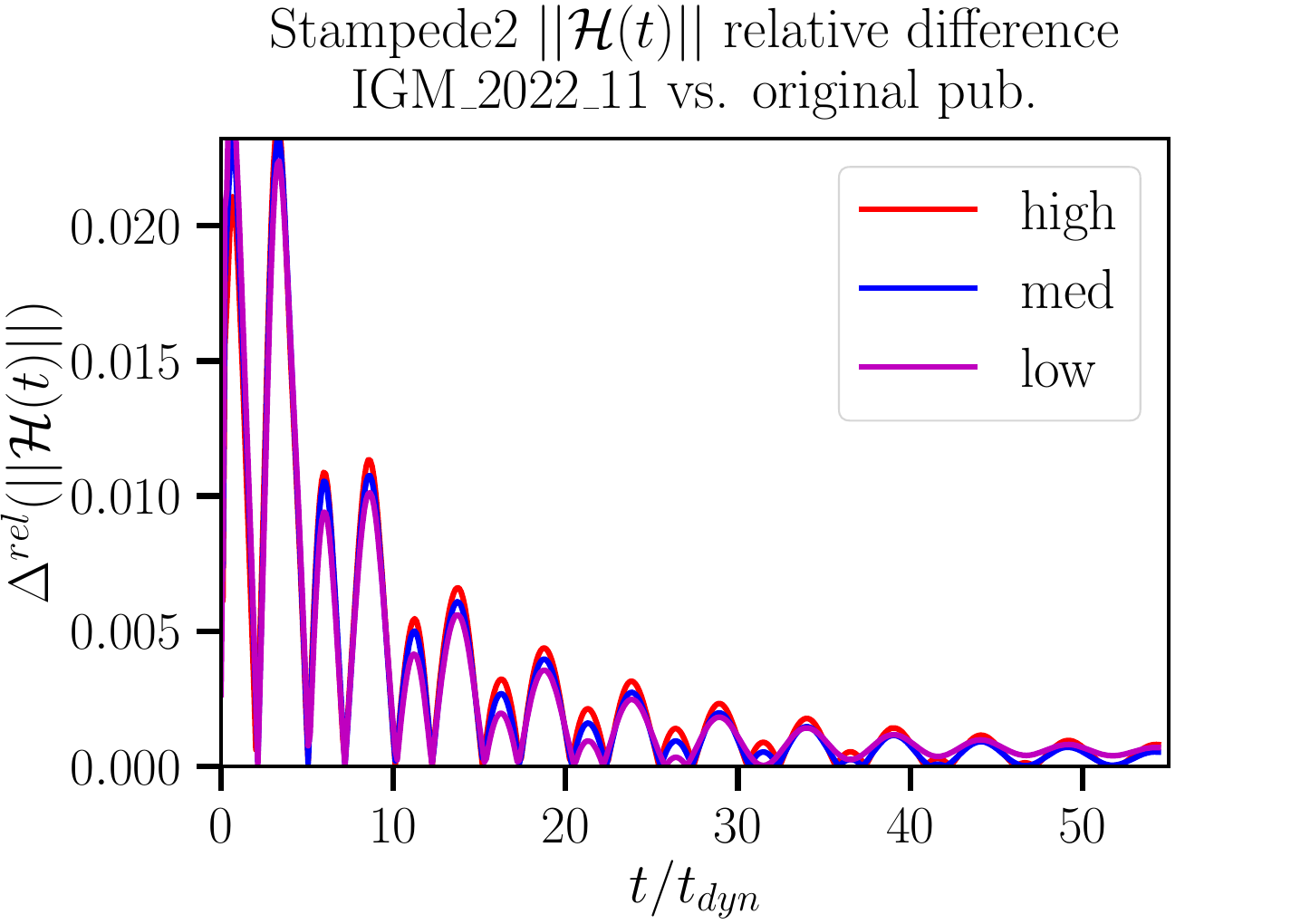}
      \caption{\ilgrmhd 2022}
    \end{subfigure}%
    \caption{Comparing the Hamiltonian constraint $\|\mathcal{H}\|$ for simulations created with two versions of \etk on Stampede2. This is the same comparison as in figure \ref{fig:expanse_ham}}
    \label{fig:stampede2ham}
\end{figure}

Figures~\ref{fig:stampede2rho} and \ref{fig:stampede2ham} display equivalent 
results for $\Delta \rho_c$ and $\|\mathcal{H}\|$ obtained on Stampede2. The same effects as observed on Expanse are evident, including differences caused by changes in the source code between \ilgrmhd 2015 and \ilgrmhd 2022

\begin{figure}
    
\end{figure}

\subsubsection{Simulation result comparison between Stampede2 and Expanse}\label{subsubsec:comparison}

\begin{figure}
    \begin{subfigure}{.55\textwidth}
      \centering
      \includegraphics[width=.95\linewidth]{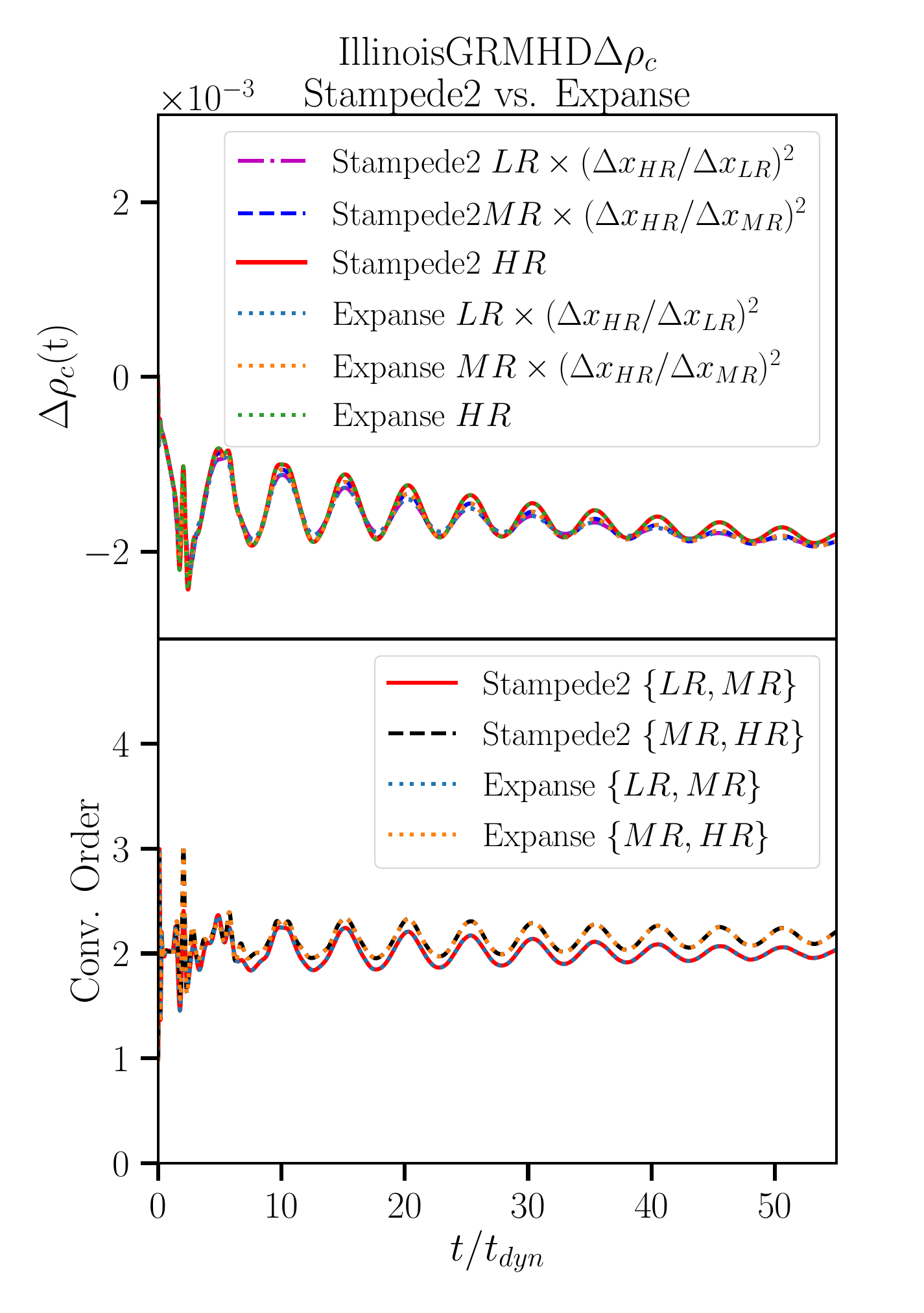}
      \label{fig:sfig1}
    \end{subfigure}%
    \begin{subfigure}{.55\textwidth}
      \centering
      \includegraphics[width=.95\linewidth]{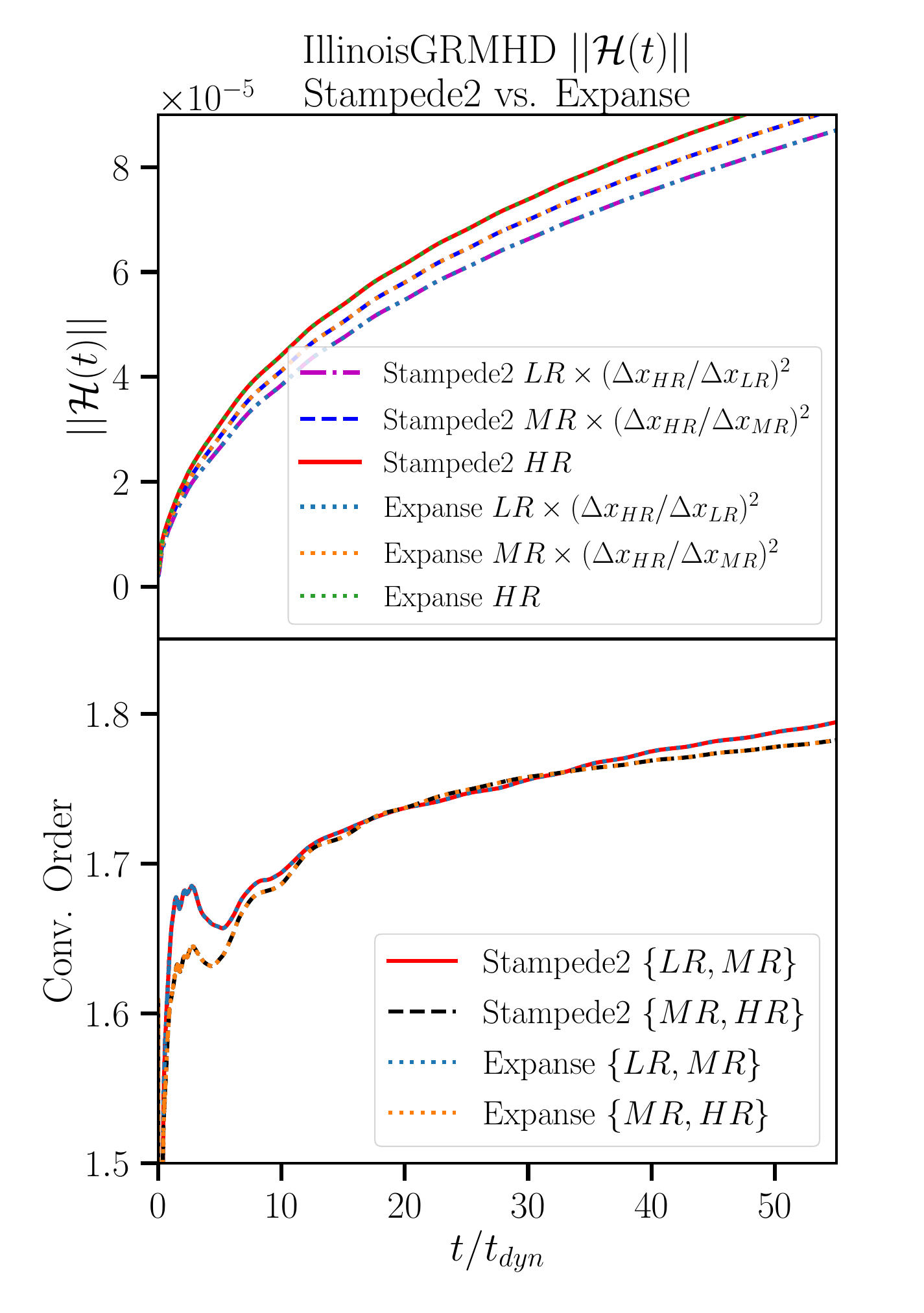}
      \label{fig:sfig1}
    \end{subfigure}%
    \newline
    \begin{subfigure}{.55\textwidth}
      \centering
      \includegraphics[width=.95\linewidth]{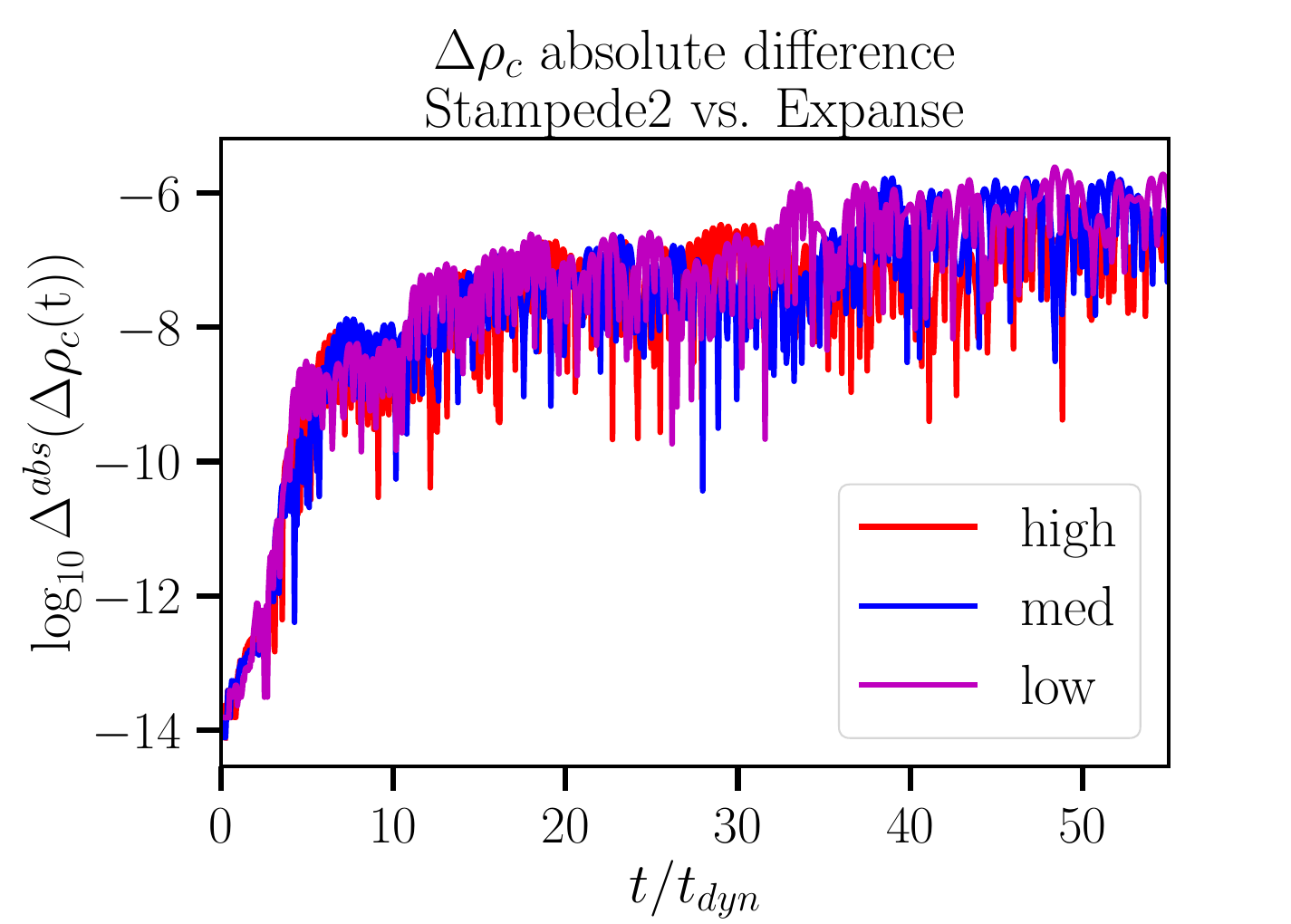}
      \caption{$\Delta \rho_c$}
      \label{fig:sfig1}
    \end{subfigure}%
    \begin{subfigure}{.55\textwidth}
      \centering
      \includegraphics[width=.95\linewidth]{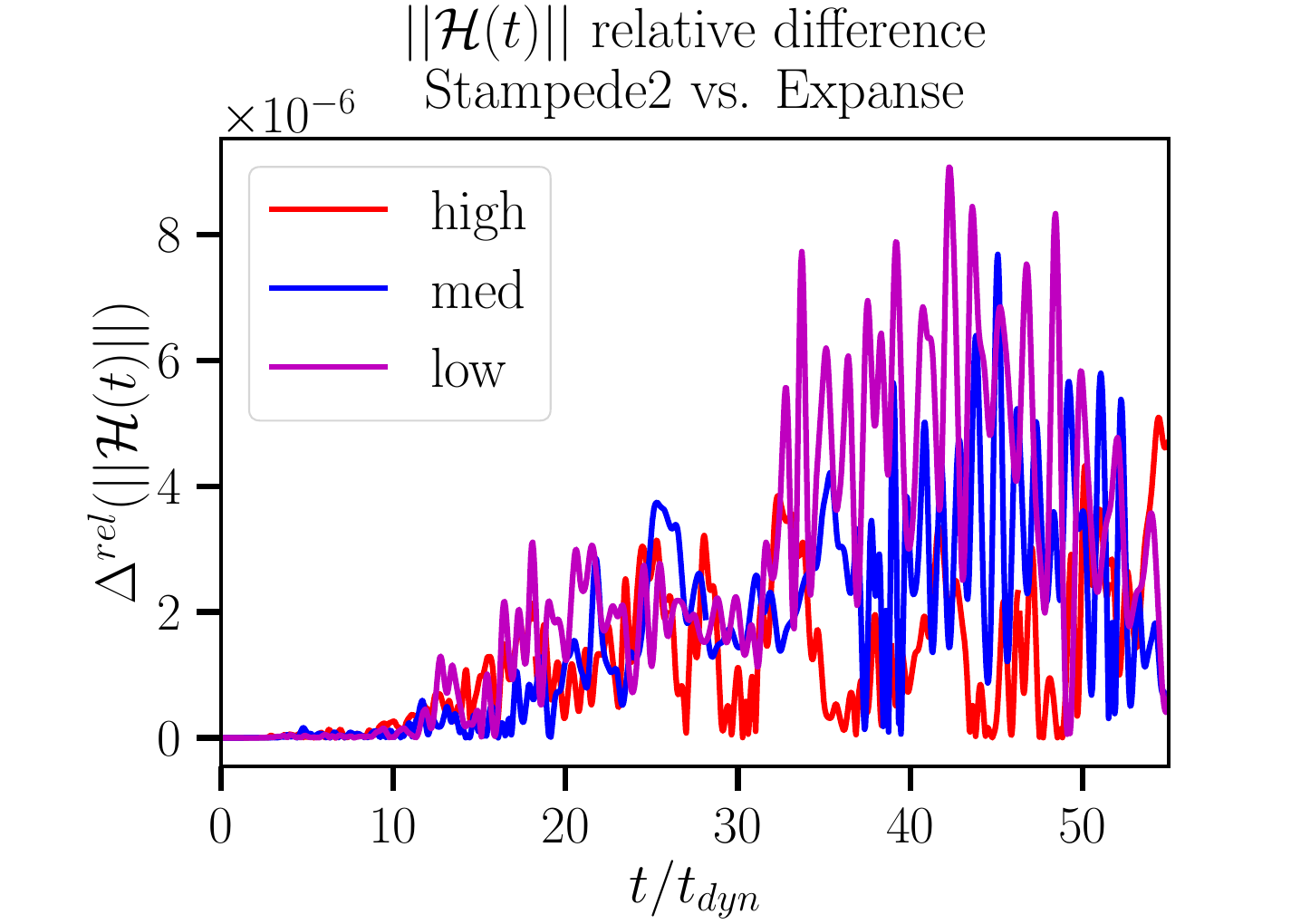}
      \caption{$\|\mathcal{H}\|$}
      \label{fig:sfig1}
    \end{subfigure}%
    \caption{The comparison of $\Delta \rho_c$ and $\|\mathcal{H}\|$ on Expanse and Stampede2. Both simulations are run using \ilgrmhd2022.}
    \label{fig:comparison}
\end{figure}

Figure~\ref{fig:comparison} compares results obtained using \ilgrmhd2022 on
Expanse and Stampede2, respectively. For both $\Delta \rho_c$ and $\|\mathcal{H}\|$ 
results agree well within the threshold for round-off agreement. The overall
results between the clusters are very similar to the results displayed in the 
left-hand columns of Figures~\ref{fig:expanse_rho}, \ref{fig:expanse_ham}, 
\ref{fig:stampede2rho}, and \ref{fig:stampede2ham} illustrating that simulations
using identical source code for \ilgrmhd{} are in round-off level agreement
both when using the code in \ilgrmhdpaper[], and \ilgrmhd2022.

\section{Conclusions}

In this study, we explore two aspects of reproducibility of computational research:
\begin{enumerate}
    \item different computing clusters using the same simulation code,
    \item same computing cluster with different versions of the simulation code.
\end{enumerate}
This poses challenges unique to scientific software, which is expected 
to compile and perform under a variety of runtime environments which are not 
known in advance.
For example, 
different compute clusters employ different compilers and different compiler
versions optimized for the cluster, some of which may fail to compile the
scientific software without modification. 
In general application software, this issue is addressed by packaging all
dependencies with the software, for example, in the form of container images as
used by the Ubuntu SNAP format~\cite{snapcraft:web}.
Containers are increasingly available on HPC systems~\cite{shifter:cug2015,10.1371/journal.pone.0177459}
providing a way to encapsulate all code dependencies except the operating system
kernel with the science code. This, however, typically entails a loss of
efficiency that may be unacceptable for high-performance
computing~\cite{P3HPC9309042, CANOPIE8950982}, and compiling from source remains
the norm for HPC codes in the current state of scientific software.

With this caveat, we observed that the same simulation code produced results agreeing to the round-off level.
Our simulations using \ilgrmhd 2015 agree with the original 
publication results on both Hamiltonian constraint and central density drift to 
round-off level, as shown in the lower left 
panel of figures~\ref{fig:expanse_rho}, \ref{fig:expanse_ham}, \ref{fig:stampede2rho}, and \ref{fig:stampede2ham}.

During our experimentation with IGM 2015, we identified two minor discrepancies in \ilgrmhdpaper[]. The first discrepancy concerns the description of the perturbation used in Section 4.1; it is accurately described in Section 4.2 of \ilgrmhdpaper but incorrectly presented in Section 4.1. Specifically, the factor $\epsilon$ of the perturbation $1 + \epsilon$ is a random number generated within the range $[0, 10^{-15})$, contrary to the fixed number $10^{-15}$ stated in Section 4.1 of \ilgrmhdpaper[]. Second, our simulations using \ilgrmhd 2022 showed discrepancies from the \ilgrmhdresult~as we stated in section \ref{sec:result_expanse}. We tracked down to a change in the source code, and reverting this change restored the reproducibility of \ilgrmhdresult.

Computational reproducibility is essential to the continuing development of 
scientific software, especially when a new module or functionality is added. In 
addition, it is also critical for the use of scientific code by others and the 
verification of results by others.

Furthermore, long-term computational reproducibility is an important 
aspect researchers be aware of. New researchers joining the field should be 
able to track and understand how to reproduce and interpret existing numerical 
experiment results. With this in mind, we 
suggest best practices for manuscripts announcing new scientific software 
or simulation results. 
\begin{itemize}
    \item Creating a DOI for specific versions of source code or depositing the source code used in the paper. To help the scientific community reproduce the results, it is recommended that papers include a trackable identifier such as DOI or Git Hash. With these unique identifiers, the simulation results and the software itself can be understood. 
    \item Attaching parameter files as supplemental materials to the paper. If the parameter files cannot be submitted as supplemental material, authors should create a DOI for parameter files used to run simulations that produce results claimed in the paper. Due to newer versions of the Einstein Toolkit, we found that we have to change some parameters of the simulation parameter files to reproduce the results claimed in \ilgrmhdpaper[]. Along with a trackable version of the simulation software, a parameter file associated with a DOI makes the simulation easy to reproduce, and the simulation workflow transparent. 
    \item Separate the computational and physical discussion. Authors for a paper with numerical experiment results should have separate sections to discuss the physical result and computational results, and how to reproduce them. Theories and problem setups should be discussed in the physical result part. Numerical experiment setups, such as simulation framework and parameter files should be introduced in the computational result section. 
\end{itemize}

In conclusion, this work studied the challenges encountered when reproducing scientific results
obtained with a state-of-the-art, real-world science code when applied to a
common test problem. This test problem was originally used to verify both code
correctness and performance in \ilgrmhdpaper[].

Science codes, and compilers used in high-performance computing
architectures, are not designed to provide bit-identical results despite identical
source code and input, instead allowing for deviations to some extent between results as
long as those
deviations are considered ``small'' compared to intrinsic approximation and 
discretization errors of the method used. 

Additionally, scientific codes are undergoing
constant changes as bugs are discovered and fixed and new features are added to them
to study new phenomena. These changes result in numeric values differing at
levels above round-off error deviations. Often these fixes are
only documented in a revision control system and not in an explicit change log
file. A key task of determining reproducibility is thus to identify these value
changing bug-fixes and quantify whether the observed differences are compatible
with the change introduced to the code.
Hitherto, this process requires an expert
understanding of the science code and is not automated. 

We demonstrate that, within  these constraints, results obtained using
\ilgrmhd 2015 can
be independently reproduced on multiple clusters and with multiple versions of
\code{IllinoisGRMHD}. 
We also provide suggestions to remedy some computational challenges encountered in this reproducibility study.

\section{Future work}

In this work, we have not attempted to extend the notion of reproducibility
past value changing bug fixes in the code. Some codes, for
example~\cite{spec:web}, attempt to address this
issue by explicitly marking commits that introduce single-bit changes in
results while others~\cite{EinsteinToolkit:2021_05} record a level of
fuzziness within which
results are considered equal. Neither approach seems fully satisfactory
and a more robust method based on the notion of equal up to round-off error may
help provide a better handle on code changes that affect reproducing scientific
results. 

\ack
Y.L. and R.H. acknowledge support from the National Science Foundation, Office of Advanced Cyberinfrastructure (OAC) through Award Number 2004879. Y.L. was also partially supported by the U.S. Department of Energy, Office of Science, Office of High Energy Physics, under Award Number DE-SC0019022. Y.L. thanks Chad Hutchens for assistance in depositing the dataset from this paper. We thank two anonymous referees for their helpful and detailed comments.

\section*{Data availability statement}

The data that support the findings of this study are openly available in the WyoScholar data repository at \wyodoi.

\section*{Code availability statement}

The simulation codes used in this study are openly available. \ilgrmhd 2015 is available at \href{https://doi.org/10.5281/zenodo.7545717}{doi:10.5281/zenodo.7545717}. \ilgrmhd 2022, which is part of Einstein Toolkit \code{ET\_2022\_11} ``Sophie Kowalevski'' release, is available at \href{https://doi.org/10.5281/zenodo.7245853}{doi:10.5281/zenodo.7245853}. The data analysis and visualization codes used to generate figures in this manuscript are publicly available at \wyodoi.

\appendix
\section{Steps to compile the codes}\label{sec:compilecode}
In the following section, we provide detailed instructions on how to compile the 
codes, \ilgrmhd 2015 and \ilgrmhd 2022, on the two 
clusters, Expanse and Stampede2.
\ilgrmhd 2015 is the code used in \ilgrmhdpaper available from its author's website and 
from \cite{zachariah_etienne_2022_7545717}. \ilgrmhd 2022 is the ``Sophie Kowalevski'' 
(\verb|ET_2022_11|) release of the Einstein Toolkit available 
from~\cite{EinsteinToolkit:web}.
To compile \ilgrmhd 2022 on either Expanse or Stampede2-skx, the steps are documented on the Einstein Toolkit website~\cite{EinsteinToolkit:web}
\begin{lstlisting}[language=sh, basicstyle=\footnotesize, caption={Compiling \ilgrmhd 2022}, literate ={-}{-}1]
curl -kLO \
  https://raw.githubusercontent.com/gridaphobe/CRL/ET_2022_11/GetComponents
chmod a+x GetComponents
./GetComponents --parallel \
  https://bitbucket.org/einsteintoolkit/manifest/raw/ET_2022_11/\
einsteintoolkit.th

cd Cactus
./simfactory/bin/sim setup-silent
./simfactory/bin/sim build --thornlist thornlists/einsteintoolkit.th
\end{lstlisting}
is sufficient.
Compiling \ilgrmhd 2015 on the other hand, due to changes in compilers and cluster environment, requires some additional steps.

We begin by downloading the code from~\cite{zachariah_etienne_2022_7545717}:
\begin{lstlisting}[language=sh, basicstyle=\footnotesize, caption={\ilgrmhd 2015: step 1}]
curl -OL 'https://zenodo.org/record/7545717/files/IllinoisGRMHD_'\
'Sept_1_2015_public_release__based_on_ET_2015_05.tar.gz'

tar xf IllinoisGRMHD_Sept_1_2015_public_release__based_on_ET_2015_05.tar.gz
cd IllinoisGRMHD_Sept_1_2015_public_release__based_on_ET_2015_05
\end{lstlisting}
Next we replace \code{Simfactory2015} with \code{Simfactory2022} to use its machine definition files
\begin{lstlisting}[language=sh, basicstyle=\footnotesize, caption={\ilgrmhd 2015: step 2}]
 rm -r simfactory
 cp -a ../Cactus/simfactory/ simfactory
\end{lstlisting}
Step 3: \ilgrmhd 2015 contains legacy code that GCC version 10 and newer flags as invalid and requires adding \verb|-fallow-argument-mismatch| to \verb|F90FLAGS| and \verb|F77FLAGS| variables, and \verb|-fcommon| to \verb|CFLAGS| and \verb|CXXFLAGS| variables in \verb|simfactory/mdb/optionlists/expanse-gnu.cfg|.
Step 4: \code{Simfactory2022} no longer provides \verb|F77|, \verb|F77FLAGS|, \verb|F77_DEBUG_FLAGS|, \verb|F77_OPTIMISE_FLAGS|, \verb|F77_NO_OPTIMISE_FLAGS|, \verb|F77_PROFILE_FLAGS|, \verb|F77_OPENMP_FLAGS|, and \verb|F77_WARN_FLAGS| and instead uses \verb|F90| etc. throughout. To compile \ilgrmhd 2015 we duplicate all \verb|F90| settings and rename the copies to \verb|F77| settings in \verb|simfactory/mdb/optionlists/stampede2-skx.cfg|.
Intel compilers version 17 or higher contain an apparent compiler bug at high optimization levels that makes them fail to compile file \verb|bbox.cc|. Thus we add
\begin{lstlisting}[language=c, basicstyle=\footnotesize, caption={\ilgrmhd 2015: step 5}]
    #if __INTEL_COMPILER >= 1700
    #pragma GCC optimization_level 1
    #endif
\end{lstlisting}
at the top of \verb|repos/carpet/CarpetLib/src/bbox.cc|.
With these modifications in place we compile using the \verb|ThornList| file supplied in the main directory of \ilgrmhd 2015:
\begin{lstlisting}[language=sh, basicstyle=\footnotesize, caption={\ilgrmhd 2015: step 6}, literate ={-}{-}1]
cd IllinoisGRMHD_Sept_1_2015_public_release__based_on_ET_2015_05
./simfactory/bin/sim setup-silent
./simfactory/bin/sim build --thornlist ThornList
\end{lstlisting}

\section{Steps to start the simulations}\label{sec:startsimulations}
We use a modified version of the file \verb|exe/tov_star_parfile_for_IllinoisGRMHD.par| included in \ilgrmhd 2015. The included parameter file is set up for a short validity test and requires some modification to match the file used in~\cite{Etienne:2015cea}.
Step 1: we change the value of \verb|CoordBase::dx|, \verb|CoordBase::dy|, and \verb|CoordBase::dz| from $1.0$ to $0.5$, $0.4$, and $0.32$ for low-resolution, medium-resolution, and high-resolution simulations, respectively. \\
Step 2: we enable the \verb|time| termination condition and set the final time to $155$
\begin{lstlisting}[language=perl, basicstyle=\footnotesize, caption={Final time}]
#cactus::cctk_itlast = 128
Cactus::terminate           = "time"
Cactus::cctk_final_time     = 155
\end{lstlisting}
Step 3: for convenience of output we change \verb|IO::out_dir|'s value to \verb|$parfile|.\\
Step 4: finally we add checkpointing and recovery options at the end of the file:
\begin{lstlisting}[language=perl, basicstyle=\footnotesize, caption={Termination and checkpoint}]
ActiveThorns = TerminationTrigger
TerminationTrigger::max_walltime                 = @WALLTIME_HOURS@
# Trigger termination 30 minutes before the walltime is reached
TerminationTrigger::on_remaining_walltime        = 30
TerminationTrigger::output_remtime_every_minutes = 30
TerminationTrigger::termination_from_file        = yes
TerminationTrigger::termination_file             = "terminate.txt"
TerminationTrigger::create_termination_file      = yes
CarpetIOHDF5::checkpoint                  = yes
IO::recover                                 = "autoprobe"
IO::checkpoint_on_terminate                 = yes
\end{lstlisting}

\section{Steps to compute expected round-off level differences}\label{sec:roundoffdifferences}

We estimate the effect of round-off level changes induced, e.g., by compiler
applied code optimization on the time evolution of results by explicitly adding
a small, random perturbation of $10^{-15}$ relative size to the initial data 
for all primitive variables. Comparing the results of this perturbed initial data with an
unperturbed simulation provides an estimate for these effects and establishes 
an order of magnitude estimate for which differences are compatible with 
round-off level changes in the data.

\ilgrmhd 2022 already contains  facilities to add such a perturbation
in the \code {ID\_converter\_ILGRMHD} module:
\begin{lstlisting}[language=perl, basicstyle=\footnotesize, caption={Random perturbation}]
ID_converter_ILGRMHD::random_pert = 1e-15
\end{lstlisting}
\ilgrmhd 2015 on the other hand contains a bug that renders the
parameter ineffectual and we apply the git commit hash 
\code{f822e2278695615a9ad508d58fe25b0c94451a31} 
``WVUThorns/ID\_converter\_ILGRMHD: When adding an optional perturbation to 
the initial data, the perturbation should be applied to all IllinoisGRMHD 
quantities, not HydroBase, at this part of the routine. Behavior was correct 
except for density. This one-line patch fixes that.'' from 
\ilgrmhd 2022 which fixes the bug.

\section*{References}
\bibliographystyle{iopart-num}
\bibliography{references}

\end{document}